\documentclass{aa}  
\usepackage{graphicx}
\usepackage{txfonts}
\usepackage{amssymb}
\usepackage{amsmath}
\usepackage{savesym}
\savesymbol{tablenum}
\usepackage{siunitx}

\newcommand{\PD}[2]{\frac{\partial #1}{\partial #2}}

\newcommand{\bvf}{Brunt--V\"ais\"al\"a}

\newcommand{\micromu}{\si{\micro Hz}}

\newcommand{\msol}{\ensuremath{\mathrm{M}_\odot}}

\begin{document}

   \title{Internal gravity waves in massive stars}

   \subtitle{II. Frequency analysis across stellar mass}

   \author{R. P. Ratnasingam
          \inst{1}
          \and
          T. M. Rogers\inst{2}
          \and
          S. Chowdhury\inst{1}
          \and
          G. Handler\inst{1}
          \and
          R. Vanon\inst{2}
          \and
          A. Varghese\inst{2}
          \and
          P. V. F. Edelmann\inst{3}
          }

   \institute{Nicolaus Copernicus Astronomical Center, Bartycka 18, 00-716              Warszawa, Poland \\
              \email{rathish.ratnasingam@newcastle.ac.uk}
         \and
             School of Mathematics, Statistics and Physics, Newcastle University, UK \\
         \and
             Computer, Computational and Statistical Sciences (CCS) Division and Center for Theoretical Astrophysics (CTA), Los Alamos National Laboratory,  Los Alamos, NM
        87545, USA
             }

   \date{}

% \abstract{}{}{}{}{} 
% 5 {} token are mandatory
\abstract{Stars that are over 1.6 solar masses are generally known to possess convective cores and radiative envelopes, which allows for the propagation of outwardly travelling internal gravity waves (IGWs). Here, we  study the generation and propagation of IGWs in such stars using two-dimensional, fully non-linear hydrodynamical simulations with realistic stellar reference states from the one-dimensional stellar evolution code, Modules for Stellar Astrophysics. Compared to previous similar works, this study utilises radius-dependent thermal diffusivity profiles for five different stellar masses at the middle of the main sequence: 3\msol{}, 5\msol{}, 7\msol{}, 10\msol{}, and 13\msol{}. From the simulations, we find that the surface perturbations are larger for higher masses, but no noticeable trends are observed for the frequency slopes with different stellar masses. The slopes are also similar to the results from previous works. We compared our simulation results with stellar photometric data from a recent survey and we found that for frequency intervals above 8 \micromu{}, there is a good agreement between the temperature frequency slopes from the simulations and the surface brightness variations of these observed stars. This indicates that the brightness variations are caused by core-generated IGWs.} 

   \keywords{Hydrodynamics --
                Waves --
                Stars: massive
               }

   \maketitle
%
%-------------------------------------------------------------------
\section{Introduction} \label{sec:intro}
Given the progress made in the field of asteroseismology (the study of pulsations and oscillations in stars apart from the Sun), it is clearly an auspicious time for theoretical research on stellar interiors to take the centre stage. Direct observables from stars, such as brightness variations from photometric data and stellar light spectrum variations from spectroscopic data, can be analysed further with asteroseismological techniques to quantify fluid oscillations in stellar interiors. These oscillations take the form of p-modes (pressure modes) and g-modes (gravity modes). Fortunately, hydrodynamical simulations can provide information on how these g-modes and p-modes are formed and how they propagate through a star.  With a rapidly increasing sample of stars available for asteroseismology from space missions, there is a strong call for studies of the effect of stellar mass and age on surface observables. 

Thanks to helioseismology (the study of pulsations and oscillations in the Sun) providing a huge amount of information on solar-like stars, we turn to more massive stars, which have not been studied as much. One particular characteristic of more massive stars is the presence of a convective core below a layer of radiative envelope, instead of a radiative core encompassed by a convective envelope in less massive stars (e.g.\ the Sun). The radiative layer is always stably stratified and subadiabatic, which allows for the propagation of outwardly travelling internal gravity waves (IGWs), generated mostly at the radiative-convective boundary. 

It is important to study these waves in massive stars, as they provide information about the top of the convective cores. Currently, one of the more important unanswered questions in this field  is the role of convection in generating these waves. \cite{1999ApJ...520..859K} suggests that the bulk material inside the convection zone generates waves, which then tunnel through and manifest themselves at the top of the zone. This notion was developed further in \cite{2013MNRAS.430.2363L} and a theoretical spectrum with a cut-off point at the convective turnover frequency was introduced. A more numerical approach was done in \cite{2013ApJ...772...21R} and the results from the two-dimensional (2D) equatorial annulus simulation of a three-solar mass (\msol{})  star showed a shallower generation spectrum. This work was improved upon with three-dimensional (3D) simulations in \cite{philipp3dpaper}, where the authors arrived at very similar results. More recently, \cite{pincon2016a} theoretically showed how plume models can lead to observations of more than one slope in the frequency spectrum of IGWs. 

In addition to the issue of uncertainty in the generation spectrum, IGW propagation in the radiation zone is also a subject of widespread study. One of the very first studies on this was carried out by \cite{1981ApJ...245..286P}, who showed that using a locally Bousinesseq and globally anelastic assumption, the linear propagation of IGWs can be represented with an analytical model. This work took into consideration the effect of pseudo-momentum conservation, geometry, and radiative damping on IGW propagation. \cite{1997A&A...322..320Z} revisited this work and used the same expression for both viscous and radiative damping interchangeably, depending on which effect was more dominant for a given case. \cite{rathish2020} showed that the 2D and 3D linear models differ by a factor of the square root of the radial coordinate. Apart from linear analyses, non-linear simulations \citep{philipp3dpaper,Horst2020} have also been done to investigate the propagation of IGWs.

Until now, numerical simulations of a convective core underlying a radiative envelope were done for mainly a 3~M$_{\odot}$ star, as done in \cite{2012ApJ...758L...6R} and \cite{philipp3dpaper}. With the assumption that the IGW spectrum can be scaled to represent that from stars of different masses and ages, \cite{2015ApJ...806L..33A} showed that there was a match between stellar brightness variations and the IGW signatures predicted by numerical simulations. However, the generation spectrum and, hence, the surface IGW spectrum will be different for stars of different ages and masses for a range of reasons, such as stellar size, where heavier and older stars tend to be larger, and convective velocities, where heavier and older stars have larger convective velocities.

In Section 2, we present our numerical setup, which describes the numerical methods, equations, and the code used in this study. Section 3 explains the properties of the stellar interiors and how they were modified from models produced by the 1D stellar evolution code named Modules for Stellar Astrophysics or MESA\citep{mesa_1,mesa_2,mesa_3,mesa_4,mesa_5} to ensure a stable running of the hydrodynamical simulations. Section 4 describes our results from simulations of different stellar models. Section 5 compares these results with some photometric data from a recent TESS survey. The final section of the paper presents a summary and our conclusions.

\begin{table*}
\caption{Radial resolution used for all the models in the convection zone ($\Delta \mathrm{r}_{\mathrm{CZ}}$) and the radiation zone ($\Delta \mathrm{r}_{\mathrm{RZ}}$). We also include the total radius of the simulation domain in centimetres ($\mathrm{r}_\mathrm{max}$/$10^{11}$) and in units of the total stellar radius ($\mathrm{r}_\mathrm{max}$/r$_{\mathrm{star}}$). The final column represents the radius of the convection zone.}             % title of Table
\label{tab:res_conv_models}      % is used to refer this table in the text
\centering                          % used for centering table
\begin{tabular}{cccccc}        % centered columns (4 columns)
\hline\hline                 % inserts double horizontal lines
Model & $\Delta \mathrm{r}_{\mathrm{CZ}}$ /$10^7$ cm & $\Delta \mathrm{r}_{\mathrm{RZ}}$ /$10^7$ cm & $\mathrm{r}_\mathrm{max}$/$10^{11}$ cm & $\mathrm{r}_\mathrm{max}$/r$_{\mathrm{star}}$ & $R_{CZ}$/$10^{11}$ cm\\
                        \hline
                        3 \msol{} midMs & 3.68 & 18.16 & 2.01 & 0.9 & 0.19\\
                    5 \msol{} midMs & 5.92 & 23.77 & 2.68 & 0.9 & 0.30\\
                    7 \msol{} midMs & 7.99 & 27.95 & 3.21 & 0.9 & 0.42\\
                        10 \msol{} midMs & 11.10 & 33.22 & 3.89 & 0.9 & 0.56\\
                        13 \msol{} midMs & 14.32 & 38.04 & 4.54 & 0.9 & 0.73\\
                        %20 \msol{} midMs & 21.32 & 58.30 & 5.86 & 0.9 & 1.10\\
\hline                                   %inserts single line
\end{tabular}
\end{table*}
\begin{figure}[ht!]
        \centering
        \includegraphics[trim={0.0cm 0.0cm 0 0.0cm},clip,width=1.0\columnwidth]{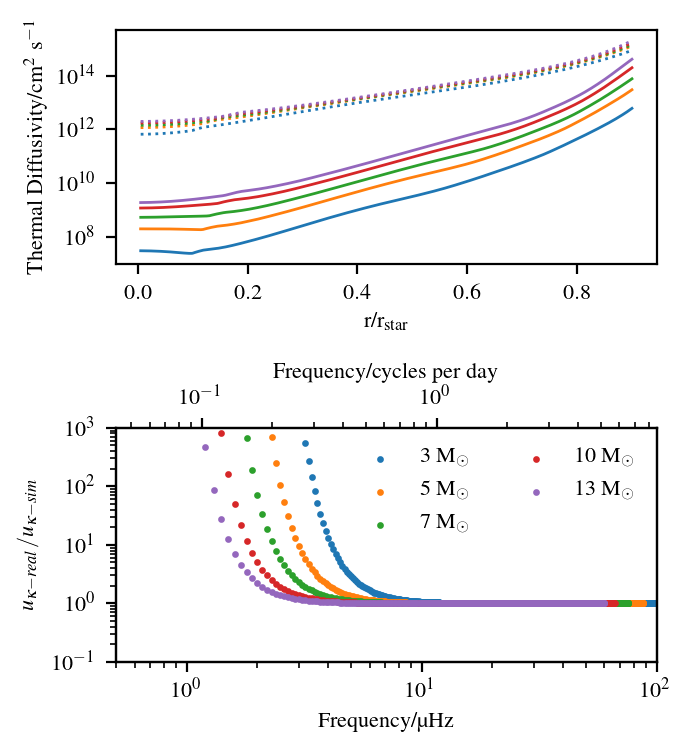}
        \caption{Thermal diffusivities (top panel) as a function of stellar radius and the ratio of linear IGW amplitudes obtained with real stellar thermal diffusivities to those obtained with $\overline{K}$ profiles used in our main simulations (bottom panel) as a function of wave frequencies at the top of the radiation zone. The dotted lines in the top panel indicate the varying thermal diffusivity profiles used for our main simulations, whilst the solid lines indicate the realistic thermal diffusivities from MESA. In the bottom panel, the wavenumbers for all the waves have been set to 1. The initial radius for these IGWs was chosen at random from the middle of the radiation zone and the initial amplitude was set to unity. \label{fig:variable_therm_diff_compare}}
        \centering
\end{figure}

\section{Numerical setup} \label{sec:Numerical_setup}
We used a pseudo-spectral method to solve the hydrodynamical momentum and energy equations, namely, Eqs.~\ref{eq:anelastic-rho} - \ref{eq:anelastic-T} in the anelastic approximation \citep{2005MNRAS.364.1135R} within the geometry of a 2D equatorial slice. An analysis of how these equations differ from the fully compressible Navier-Stokes equations and variations of  the anelastic approximation was done in \cite{Brown2012}, which referred to the following equations as the Rogers-Glatzmaier (RG) equations. These are:
\begin{align}\label{eq:anelastic-rho}
\nabla \cdot \overline{\rho} \vec{v} &= 0, \\
\label{eq:anelastic-v}
\PD{\vec{v}}{t} + (\vec{v} \cdot \nabla) \vec{v} &=
- \nabla \left(\frac{P}{\overline{\rho}}\right) - C \overline{g} \vec{r} + 2(\vec{v} \times \vec{\hat{z}} \Omega) \\
\nonumber
& + \overline{\nu} \left( \nabla^2 \vec{v} + \frac{1}{3} \nabla (\nabla \cdot \vec{v}) \right),\\
\label{eq:anelastic-T}
\PD{T}{t} + (\vec{v} \cdot \nabla) T &= (\gamma - 1) T h_\rho v_r\\
\nonumber
&- v_r \left( \frac{d\overline{T}}{dr} - (\gamma - 1) \overline{T} h_\rho \right)\\
\nonumber
& + \frac{1}{c_v\overline{\rho}} \nabla \cdot (c_p \overline{K}\overline{\rho}\nabla T) \\
&+ \frac{1}{c_v\overline{\rho}} \nabla \cdot (c_p \overline{K}\overline{\rho}\nabla \overline{T}).%\\
\end{align}
where $\rho$, $T,$ and $P$ are the perturbation density, temperature, and pressure, respectively. Any quantity with an over-line or bar represents the background state variables, which vary only with radius. The radial velocity, $v_r$, and the tangential velocity, $v_{\theta}$, together form the velocity vector, $\vec{v}$. The negative inverse density scale height is represented by $h_\rho$ = d$(\ln \overline{\rho})/$dr. The rotational angular velocity and viscosity are represented by $\Omega$ and $\overline{\nu}$. The adiabatic index, $\gamma$ is set to be 5/3. The specific heat capacity at constant volume and specific heat capacity at constant pressure are represented by $c_v$ and, $c_p$ respectively. The reference state gravity is represented by $\overline{g}$. The co-density \citep{Braginsky1995,2005MNRAS.364.1135R}, labelled as  $C$, is 
\begin{equation}
\label{eq:codensity}
C=-\frac{1}{\overline{T}}\left(T  + \frac{1}{\overline{g} \overline{\rho}}\PD{\overline{T}}{r}P \right).
\end{equation}

In the radial direction, the RG equations are solved using a uniform finite difference scheme, specifically: the central difference method (e.g. \cite{grossmann2007numerical}). In the horizontal direction, the spectral solution method was applied, where stellar variables are expanded in sines and cosines. The non-linear terms of the RG equations are integrated in time using the Adams-Bashforth method. The linear terms are integrated in time using the Crank-Nicolson method.

The temperature perturbation, $T$, and radial velocity, $v_r$, at the top boundary of the simulation domain were set to zero. For the horizontal velocities, $v_{\theta}$, at the top boundary, the following stress-free boundary condition was imposed as $\partial v_{\theta}/\partial r = 0$.

\subsection*{Setup of the convection and radiation zones}

\begin{figure}
        \centering
        \includegraphics[trim={0.0cm 0.0cm 0 0.0cm},clip,width=\columnwidth]{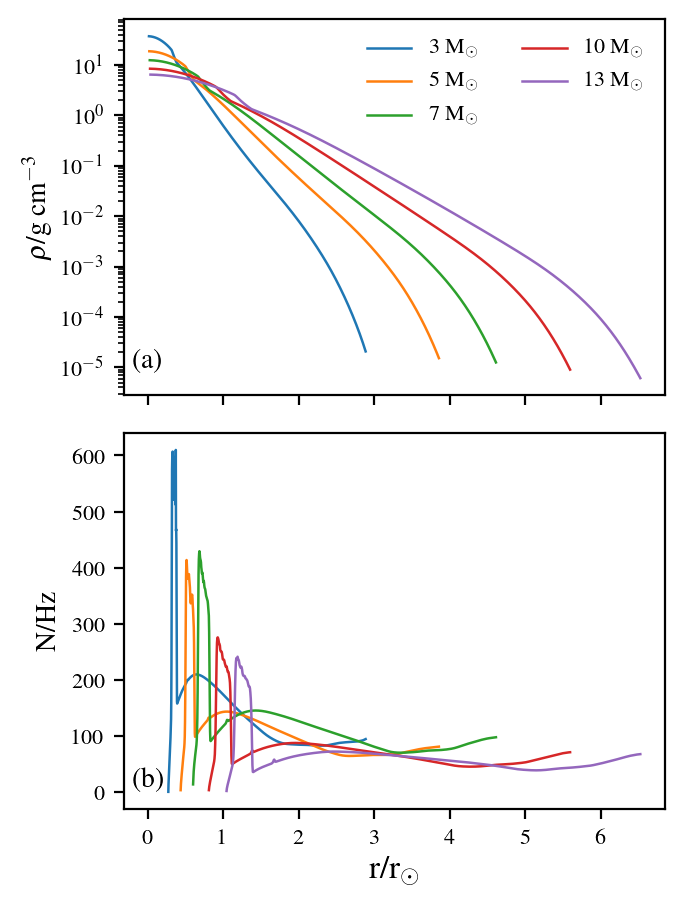}
        \caption{Density (a) and \bvf{} frequency (b) profiles as functions of stellar radius, in units of total solar radius. All the models are cut at 90\% the total stellar radius to ensure numerical stability.   \label{fig:bvf}}
        \centering
\end{figure}

The 1D stellar evolution code, MESA, was used to generate models of various stars in the intermediate-mass regime at the middle of the main sequence (midMs), when the central hydrogen mass fraction becomes approximately 0.35 These models were then set as the reference state for these simulations. The stellar metallicity, $Z$, was set to be equal to the solar value of $Z$ = 0.02. The mixing-length parameter was set to 1.8 and the convective overshoot profile was set to exponential (see Eq.~(2) in \cite{Pedersen_2018} for details). Due to limitations in the numerical solver that we utilised, the maximum density variation between the bottom and top of the simulation domain had to be capped at approximately six orders of magnitude (see Figure~\ref{fig:bvf}(a)). This was found to be within 90\% of the total stellar radius (as defined in MESA), for all of our models,  the MESA models have been cropped. We note that even with trimming, stars with larger mass have larger radii, which causes the radial resolution to decrease with increasing mass as we are limited by the total number of radial grids in our simulations. Table~\ref{tab:res_conv_models} summarises this information for all the models we simulated. 

We study intermediate-mass stars, specifically stars with a convective core and radiative envelope. In our simulations, the convection and radiation zones are set using the subadiabaticity term in Eq.~\ref{eq:anelastic-T}, as given below:
\begin{equation*}
 -\left( \frac{d\overline{T}}{dr} - (\gamma - 1) \overline{T} h_\rho \right).
\end{equation*}
This term represents the difference between the stellar temperature gradient and the adiabatic temperature gradient, where a positive value is necessary for convection and conversely, a negative value indicates a convectively stable region. It is derived with the ideal gas assumption and can be used to calculate the \bvf{} frequency using:
\begin{equation}\label{eq:RG_brunt_eq}
N^2 = \frac{\overline{g}}{\overline{T}}\left(\frac{d\overline{T}}{d\overline{r}}-\left(\gamma-1\right)\overline{T}h_{\rho}\right).
\end{equation}
However, stellar models from MESA are produced with a more realistic equation of state, which disagree with our expression for the subadiabaticity. Thus, to better represent the subadiabaticity in the radiation zone, we calculated this term from the \bvf{} frequency profile for all of our MESA models, by rearranging Eq.~\ref{eq:RG_brunt_eq}. However, in the convection zone, MESA provides negative \bvf{} frequencies which are physically inaccurate. Thus, we find a value between $1 \times 10^{-6}$ and $1 \times 10^{-7}$ that leads to the desired convective velocities once the convection in this region reaches a steady-state evolution. The realistic values of subadiabaticity in stellar convection zones have yet to be determined by the larger stellar research community. 

Figure~\ref{fig:bvf}(b) shows the \bvf{} frequencies for MESA stellar models with different masses. Generally, we see a decrease in the average \bvf{} frequency with increasing stellar mass. We also see a significant peak at the beginning of each profile, which is an artefact of the retracting convective core as a star ages. 

\begin{table*}
\caption{Stellar parameter for models used in the main simulations (see Section~\ref{sec:main_sim}) in this investigation. The average v$_{\mathrm{rms}}$ is the root-mean-square of the total velocity in the horizontal direction averaged over the radial extent of the convection zone and the time domain given in Fig.~\ref{fig:vrms_vary_kap}. The value, $K$, is a multiplication factor, which is used to in Eq.~\ref{eq:K_vary} to set the thermal diffusivity in our simulations. The viscosity, $\nu$ is a constant value throughout the stellar model. Since stellar density varies up to six orders of magnitude, from roughly $10^1$ g cm$^{-1}$ to $10^{-5}$ g cm$^{-1}$ within the models (see Fig.~\ref{fig:bvf}), the thermal diffusion varies up to three orders of magnitude (from Eq.~\ref{eq:K_vary}), and thus, the Prandtl number, $Pr = \nu / \overline{K}$, varies up to three orders of magnitude. }
\label{tab:model_param}
\centering
        \begin{tabular}{cccccc}
        \hline\hline  
                Model & $K$/10$^{12}$ cm$^{2}$ s$^{-1}$ &$\nu$/10$^{12}$ cm$^{2}$ s$^{-1}$ & $Re$ & average v$_{\mathrm{rms}}$/$10^{4}$cm s$^{-1}$ & MLT velocity/$10^{4}$cm s$^{-1}$\\
                \hline
                3\msol{} midMs & 5 & 4 & 483   & 10.5 & 0.69 $\pm$ 0.18\\
                5\msol{} midMs & 5 & 5 & 832  & 14.2 & 1.2 $\pm$ 0.19\\
                7\msol{} midMs & 5 & 5 & 1658  & 21.0 & 1.8 $\pm$ 0.42 \\
                10\msol{} midMs & 5 & 5 & 2559  & 23.6 & 2.4 $\pm$ 0.43 \\
                13\msol{} midMs & 5 & 5 & 3932  & 28.2 & 3.0 $\pm$ 0.9\\
                \hline
                %20\msol{} midMs & 8 & 8 & 3230 & 1 & 6.64 & 240000 & 50000 $\pm$ 10000 \\
        \end{tabular}
\end{table*}

\section{Properties of the stellar interior}
\subsection{Core convection}
The 1D stellar evolution code, MESA, simulates the convection zones in stars using a formulation known as mixing-length theory (MLT). Although this method is computationally inexpensive, there is much debate on the accuracy of this technique among the stellar research community. In our simulations, the average convective velocity for any particular stellar model was forced to be higher than their corresponding MLT velocity given by MESA (refer to Table~\ref{tab:model_param}). This was done to counterbalance the effect of the unrealistically large viscosities and thermal diffusivities used in our simulations, which cause stronger damping on waves than expected in a real star. Given that we are generally more concerned with wave amplitudes at the surface, the higher convective velocities allow IGWs to be better represented, as they are generated with higher amplitudes and less affected  by thermal diffusion and viscosity. Additionally, it is not certain whether the MLT velocities accurately represent convective velocities in stars \citep{Jones2017}. 

One of the key differences in core convection between different stellar masses is the magnitude of convective velocities. With increasing stellar mass, convective velocities are larger due to higher temperatures, as shown in the last column of Table~\ref{tab:model_param}, which summarises different properties of the models used in our investigation. The error values indicate the general variation of MLT velocities in the convection zone\footnote{MLT velocities approach zero close to the convective-radiative boundary, so this error sets a more accurate upper limit than a lower limit.}. Convective velocities are also expected to increase with stellar age, which will be investigated in Rogers et al.\ (in prep). Mainly due to the larger convective velocities in more massive stars, several convective parameters such as the Reynolds and Rayleigh numbers were found to be larger too, as seen in the 4th and 6th columns in Table~\ref{tab:model_param}. The Reynolds number, $Re$, is defined as 
\begin{equation}\label{eq:Re}
        Re = \frac{uL}{\nu},
\end{equation} 
where $\nu$ is the kinematic viscosity, $u$ is the characteristic velocity in the convection zone, and $L$ is the characteristic length scale.It is usually defined as the radial extent of the convection zone. This quantity describes the ratio of the inertial forces to the viscous forces in the system. Low values signify laminar flow, while high values show turbulent flow. For more massive stars, larger $Re$ indicates stronger and more turbulent convection. 

\begin{figure}[ht!]
        \centering
        \includegraphics[trim={0.0cm 0.0cm 0 0.0cm},clip,width=\columnwidth]{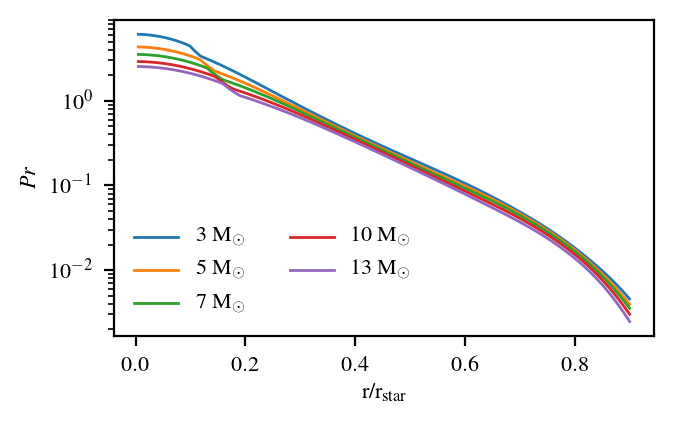}
        \caption{Prantl number profiles used in the simulations of all the stellar models used in this work. \label{fig:Pr}}
        \centering
\end{figure}

For the real intermediate-mass stars studied in this paper, we expect the average Prandtl number $Pr = \nu/\overline{K}$, to be somewhere close to 1 $\times$ 10$^{-6}$, which is much smaller than the values we use in our simulation \citep{Garaud_2015}). The work done in \cite{Garaud_2015} further shows that $Pr$ is expected to decrease with increasing stellar mass and with increasing radial distance from the stellar core within the star (Fig. 7 in \cite{Garaud_2015}). Generally, with increasing age, the stellar interior profiles of an intermediate-mass star follow closely those of higher mass stars, so we expect $Pr$ to decrease as a star ages and also with radial distance from the stellar centre. In our models, we show in Fig.~\ref{fig:Pr} that $Pr$ varies approximately from 10 at the bottom boundary to 0.01 at the top boundary. Unfortunately, this mismatch between realistic and simulated $Pr$ is a computational limitation faced by many such studies, especially when turbulent convection is involved.

\subsection{The radiation zone}
The radiation zone is a convectively stable region, where radiative diffusion represents the main form of heat transport. The stratified density in this region allows the propagation of IGWs, generated mainly in the convection-radiative boundaries. As these waves travel through the radiation zone, radiative diffusion causes these waves to be damped, which primarily depends on the thermal diffusivity and the \bvf{} frequency profiles. For linear IGWs, the effect of thermal diffusion\citep{1981ApJ...245..286P,1999ApJ...520..859K} on wave amplitudes can be written as: 
\begin{equation}\label{eq:A_exp_tau}
    A_{\mathrm{IGW}} \propto A_{0} e^{-\tau}
\end{equation}
where 
\begin{align}\label{eq:tau}
\tau = \int_{r_\text{generation}}^{r} \overline{K}\left(\frac{m^{3} N^3}{r^3 \omega^4}\right)\left(1-\frac{\omega^2}{N^2}\right)^{1/2} dr'. 
\end{align}
The damping opacity, $\tau$, is a function of radius, $r$, the wave frequency, $\omega,$ and wavenumber, $m$.

The thermal diffusivity is a function of stellar radius and increases by at least four orders of magnitude in the radiation zone (see solid lines in Fig.~\ref{fig:variable_therm_diff_compare}). However, to ensure numerical stability, we have to boost the thermal diffusivities in our simulations. Although we cannot reproduce the realistic thermal diffusivity profile in stars, we mimic the increase using the profile shown below:
\begin{equation}\label{eq:K_vary}
    \overline{K} = K\rho^{-0.5}
,\end{equation}
where $K$ is a constant, chosen to ensure numerical stability. Our $\overline{K}$ values are generally larger than in real stars except at the surface, but this is compensated by our larger convective velocities (see previous section for details). This thermal diffusivity profile was applied on five different stellar models and the kinematic viscosity was chosen to be a constant (see Table~\ref{tab:model_param} for details on the thermal diffusivity and viscosity profiles). The results of this investigation are elaborated further in Section~\ref{sec:main_sim}.

\section{Main results}\label{sec:main_sim}
To investigate the effect of our artificial thermal diffusion profile compared to the more realistic ones from MESA, we started by conducting a small study using the IGW linear theory.
\begin{figure}[ht!]
        \centering
        \includegraphics[trim={0.0cm 0.0cm 0 0.0cm},clip,width=\columnwidth]{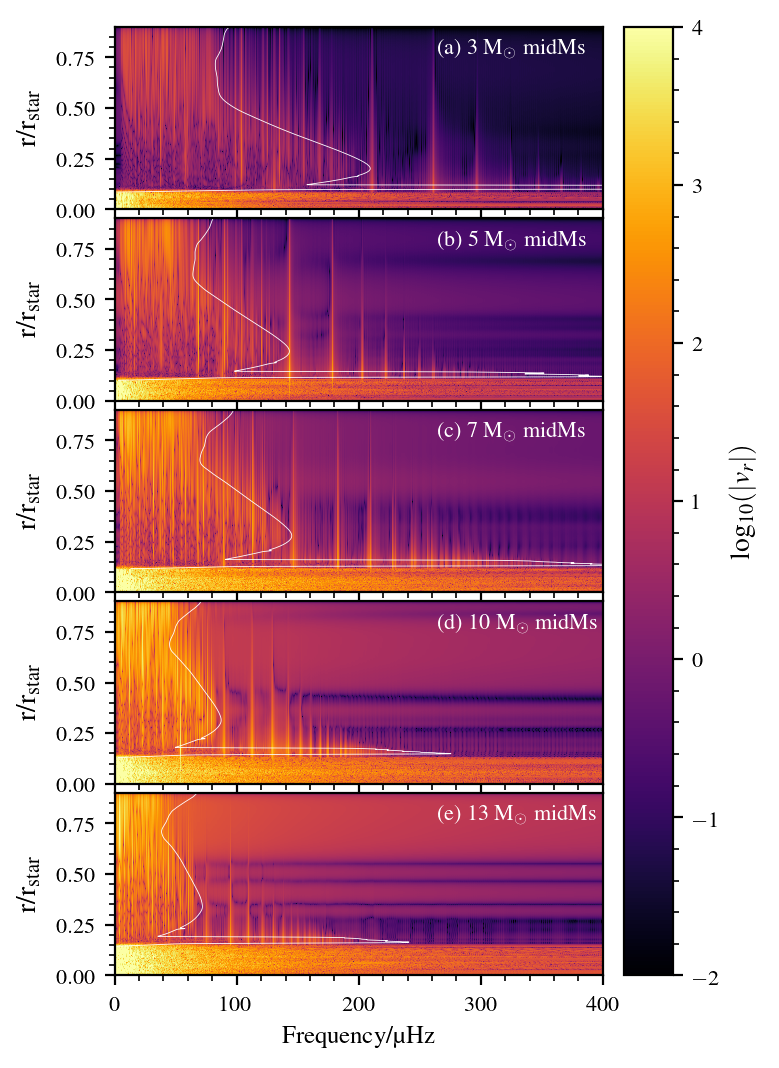}
        \caption{ Radial velocities as functions of stellar radius and frequency for different stellar masses, all at midMs. The solid, white line shows the \bvf{} frequency profile for each stellar model, respectively. \label{fig:All-Ms-midms-full-interior-freq-spectrum}}
        \centering
\end{figure}
\begin{figure}[ht!]
        \centering
        \includegraphics[trim={0.0cm 0.0cm 0 0.0cm},clip,width=0.9\columnwidth]{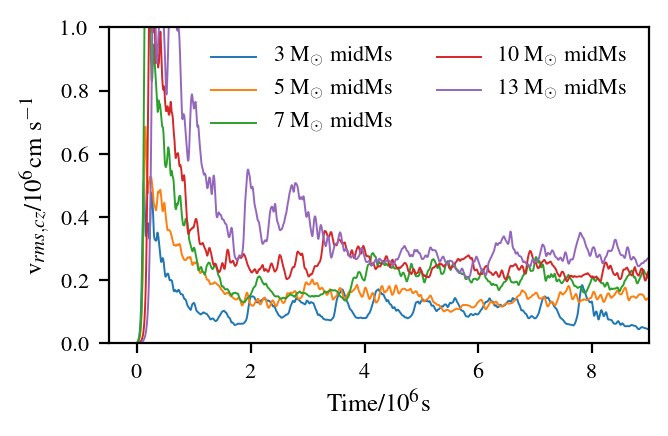}
        \caption{Average $\mathrm{v_{rms}}$ in the convection zones as a function of time of all the models in Table.~\ref{tab:model_param}. \label{fig:vrms_vary_kap}}
        \centering
\end{figure}
\begin{figure*}[ht!]
        \centering
        \includegraphics[trim={0.0cm 0.0cm 0 0.0cm},clip,width=0.8\textwidth]{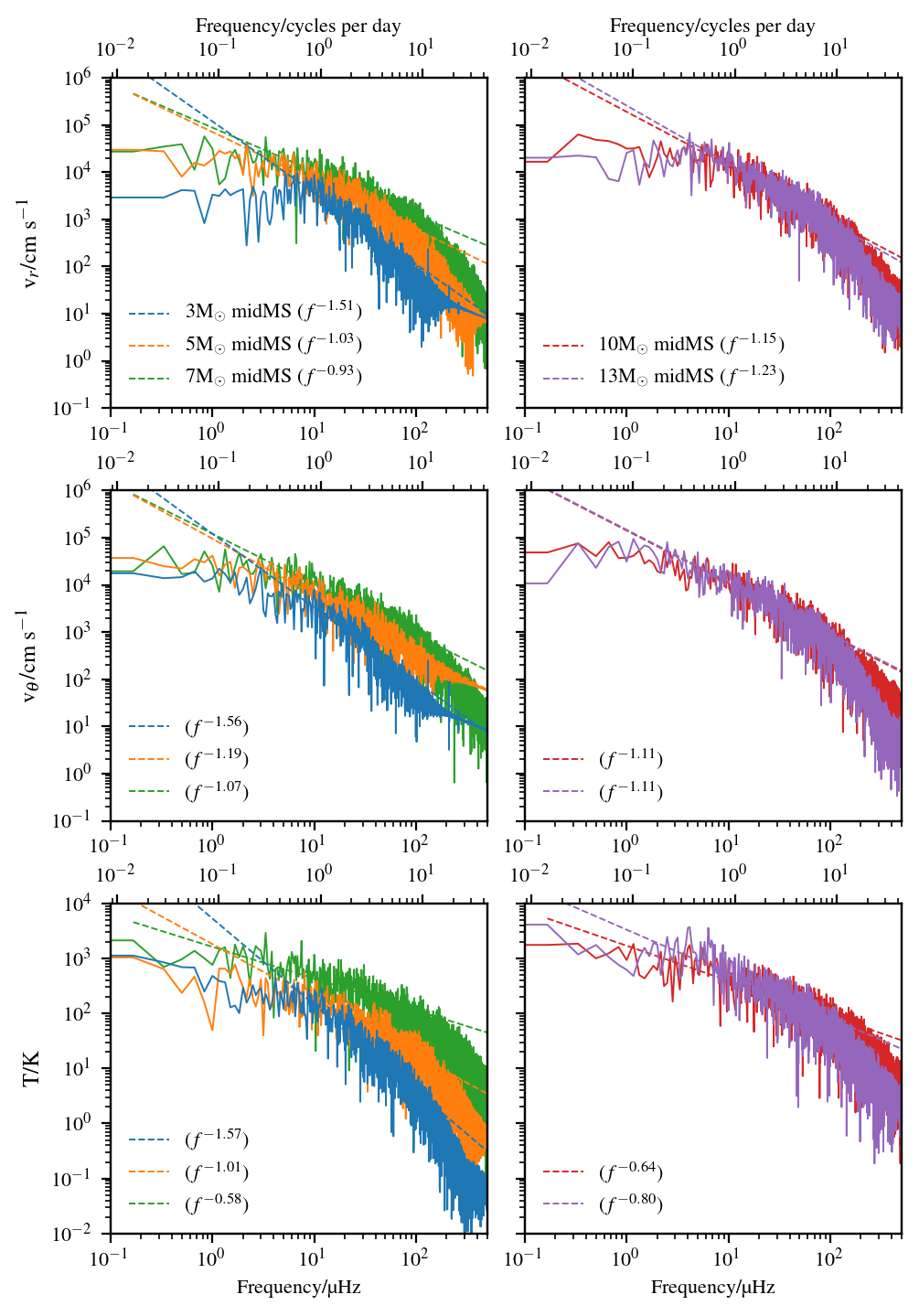}
        \caption{Radial velocities (top panels), tangential velocities (middle panels), and temperatures (bottom panels) as a function of frequency for different stellar masses in the middle of the convection zone for a randomly chosen radial ray. The dashed lines show the line fits to each profile at frequencies between 8 \micromu{} and 30 \micromu{}.  \label{fig:vtheta-vr-tem-grid200}}
        \centering
\end{figure*}
In Fig.~\ref{fig:variable_therm_diff_compare}, we show two plots, where the thermal diffusivities used in these simulations (top panel) are higher than the realistic values from MESA. From the bottom panel, which shows the ratio of linear IGW amplitudes for the thermal diffusivity from MESA to those for the artificial thermal diffusivity, we see that at lower frequencies, the artificial $\overline{K}$ predicts smaller IGW amplitudes than those from MESA's $\overline{K}$. We find that for any wave frequency above 8 \micromu{}, simulations of all the stellar models should predict wave amplitudes with higher accuracy. However, below this frequency, we see large variations in the ratios which increase with increasing stellar mass.

\begin{figure*}[ht!]
        \centering
        \includegraphics[trim={0.0cm 0.0cm 0 0.0cm},clip,width=\textwidth]{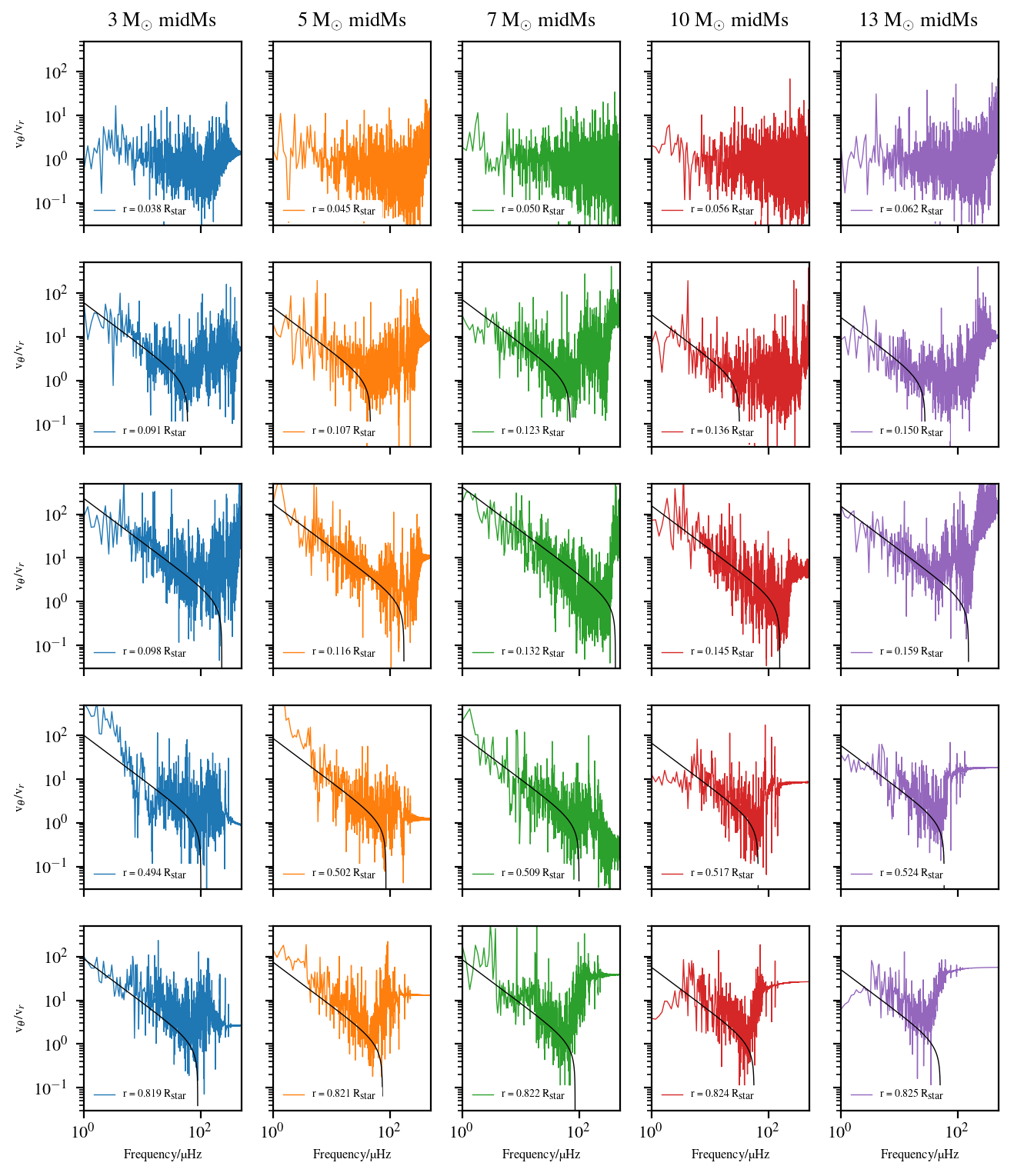}
        \caption{Ratios of the tangential velocity ($v_{\theta}$) to the radial velocity ($v_{r}$) as a function of frequency, at different radii throughout the stellar interior for different stellar masses. The black line represents the ratio of the \bvf{} frequency to the linear IGW frequency, at that particular radius. The first row shows $v_{\theta}$/$v_{r}$ inside the convection zone, the second and third rows show the ratios at 0.1 $H_P$ and 0.25 $H_P$, respectively, and the final two rows show the ratios at a randomly chosen radius inside the radiation zone and close to the top of the simulation domain, respectively.  \label{fig:ratio-vtheta-vr}}
        \centering
\end{figure*}

Moving on to the simulations, we show the average v$_{\mathrm{rms}}$ in the convection zone for all of our stellar models as a function of time in Fig.~\ref{fig:vrms_vary_kap}. Generally, we see that increasing the stellar mass leads to increasing convective velocities. Looking at individual convective velocity profiles, we see the following trend: a rapid increase in fluid velocity, indicating the onset of convection, followed by a gradual decrease and settling into a steady-state evolution. We chose a time interval of 6 $\times 10^{6}$ s (only when the convective core is in a steady-state evolution) to perform all of our time-series analysis. Figure~\ref{fig:All-Ms-midms-full-interior-freq-spectrum} shows the log of radial velocities from our simulations as a function of stellar radius and wave frequency for a randomly chosen ray or angle. All five spectra exhibit stationary modes, which are more distinctly recognisable at higher frequencies. The lower masses exhibit more modes, mainly because of the generally larger \bvf{} frequency (white, solid lines) for models with lower stellar masses. In the convection zone, velocities are larger at smaller frequencies, which shows that most of the energy in the convection zone is concentrated on larger timescale motions (e.g. 10 to 30 \micromu{}, which corresponds to approximately motions on timescales between 0.8 to 3 days). 

Taking slices of each plot in Fig.~\ref{fig:All-Ms-midms-full-interior-freq-spectrum} for all five stellar models at r = 0.5 $r_{CZ}$, where $r_{CZ}$ is the extent of the convection zone, we obtain the spectra of motions as shown in Fig.~\ref{fig:vtheta-vr-tem-grid200} (radial velocities in the top panel, tangential velocities in the middle panel, and temperatures in the bottom panel). We find two main observations, with the first one being that for all the models, there is a trend of high amplitudes at very low frequencies, which decreases rapidly at mid-range frequencies. Second, the homogenous mixing of convection has allowed the tangential and radial velocities of each stellar model to become similar  and, thus, their slopes (indicated by the dashed lines) end up becoming similar as well. To confirm this, we show the ratio of the tangential velocity to the radial velocity in the top panel of Fig.~\ref{fig:ratio-vtheta-vr}, which shows oscillatory behaviour around 1.

The panels in row 2 and 3 of Fig.~\ref{fig:ratio-vtheta-vr} show the ratio, $v_{\theta}/v_r$ at 0.1 $H_{\mathrm{P}}$ and 0.25 $H_{\mathrm{P}}$, where $H_{\mathrm{P}}$ is the pressure scale height, from the top of the convection zone, respectively. For linear IGWs, we can use the dispersion relationship for IGWs to get
\begin{equation}\label{eq:ratio_vr_vtheta_bvf}
    \frac{v_{\theta}}{v_{r}} = \left(\frac{N^2}{\omega^2} - 1\right)^{1/2}. 
\end{equation}
The term on the right in Eq.~\ref{eq:ratio_vr_vtheta_bvf} is plotted as the solid, black line in Fig.~\ref{fig:ratio-vtheta-vr}. In general, we find that the simulations agree well with the linear theory. There are a few exceptions, such as the disagreements seen between the black line and the velocities at very low frequencies for some radii and masses. At such low frequencies, our simulations do not resolve IGWs completely, so we did not include these waves in our analysis.

\begin{figure*}[ht!]
        \centering
        \includegraphics[trim={0.0cm 0.0cm 0 0.0cm},clip,width=0.8\textwidth]{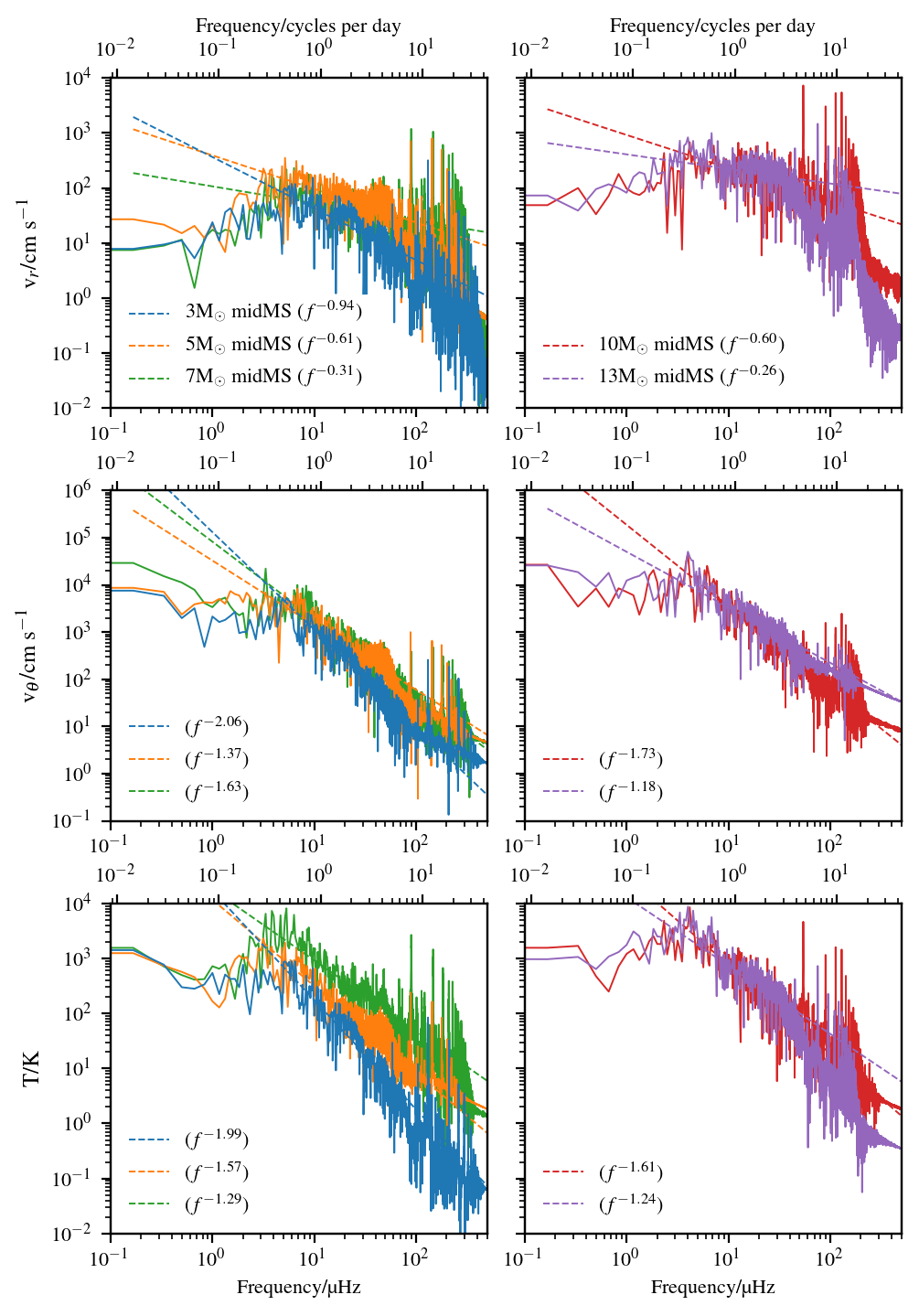}
        \caption{Radial velocities (top panels), tangential velocities (middle panels), and temperatures (bottom panels) as a function of frequency for different stellar masses at 1/4th the pressure scale height from the top of the convection zone for a randomly chosen radial ray. The dashed lines show the line fits to each profile at frequencies between 8 \micromu{} and 30 \micromu{}.  \label{fig:vtheta-vr-tem-grid513}}
        \centering
\end{figure*}
\begin{figure*}[ht!]
        \centering
        \includegraphics[trim={0.0cm 0.0cm 0 0.0cm},clip,width=0.8\textwidth]{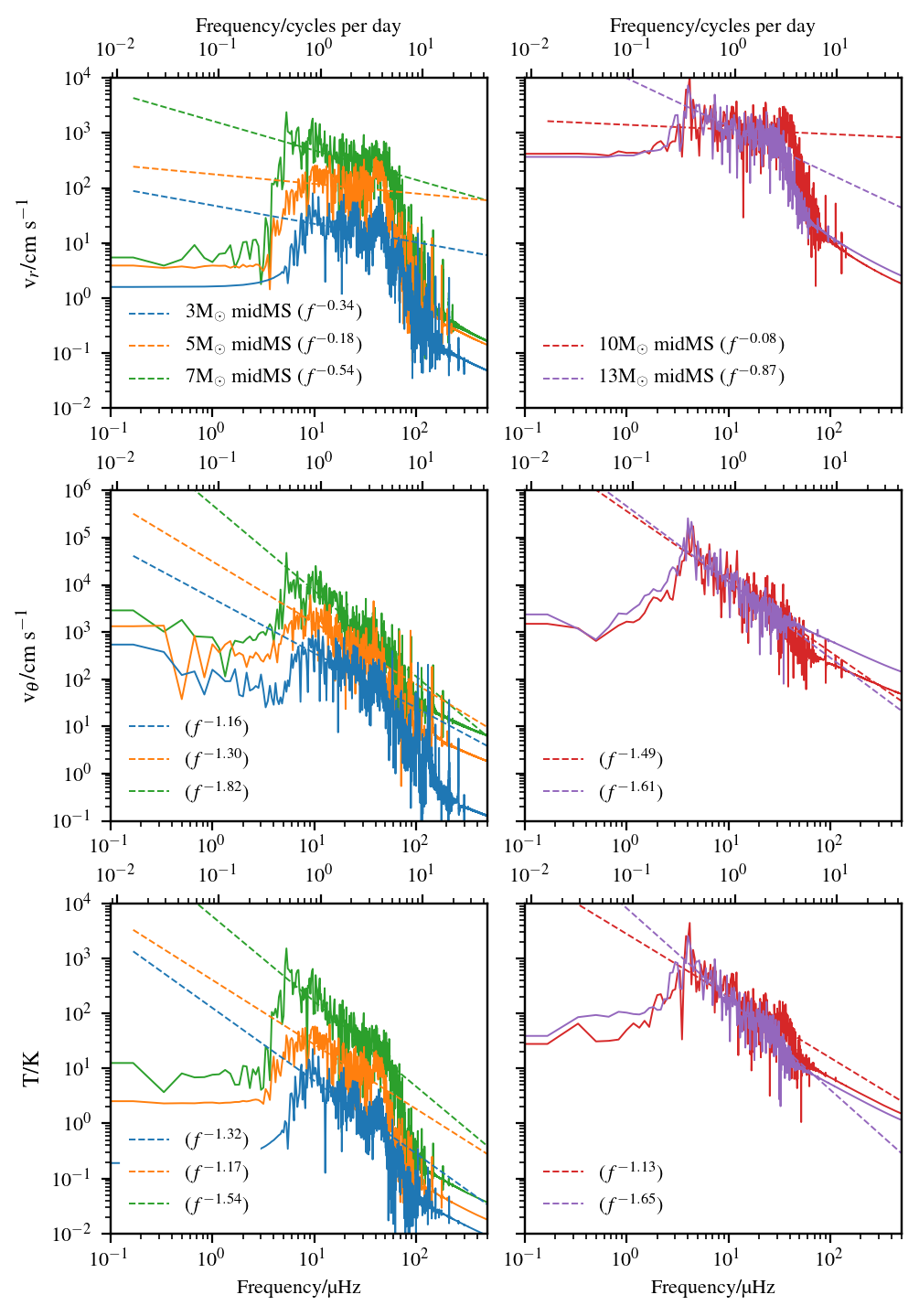}
        \caption{Radial velocities (top panels), tangential velocities (middle panels), and temperatures (bottom panels) as a function of frequency for different stellar masses at 0.82 R$_{\mathrm{star}}$ for a randomly chosen radial ray. The dashed lines are line-fits to each profile at frequencies between 8 \micromu{} and 30 \micromu{}.  \label{fig:vtheta-vr-tem-grid1400}}
        \centering
\end{figure*}

We show the velocity and temperature perturbations as a function of frequency at a randomly chosen angle and 0.25 $H_{\mathrm{P}}$ from the top of the convection zone in Fig.~\ref{fig:vtheta-vr-tem-grid513}. The radius of 0.25 $H_{\mathrm{P}}$ from the top of the convection zone chosen here are all inside the \bvf{} frequency peaks seen in Fig.~\ref{fig:bvf}(b), allowing for stationary modes at higher wavenumbers to form within this peak. Thus, we observe stationary wave peaks at higher frequencies, $f > 50$ \micromu{}, in almost all the spectra. As mentioned previously, wave amplitudes from our simulations match those calculated using linear wave theory with realistic thermal diffusivities up to 50\% accuracy above 8 \micromu{}. Thus, we chose a frequency range of 8 \micromu{} to 30 \micromu{} for the rest of the analysis. We fit these spectra with a simple straight line form, i.e. $Y = mX + C$, where $X$ is the logged frequency and Y is the logged velocity or temperature perturbation, represented by the dashed lines in Fig.~\ref{fig:vtheta-vr-tem-grid513}. The value $m$, is defined as the slope of these fits. Generally, we observe flatter slopes ($>-1$) for the radial velocities and steeper slopes for the tangential velocities and temperature perturbations ($< -1$). We also observe the temperature perturbation slopes to be more similar to the tangential velocity slopes, compared to the radial velocity slopes. 

To further check whether these motions are waves, we can manipulate the following linear, asymptotic relation for IGWs,
\begin{equation}\label{eq:ratio_vr_vtheta}
     \frac{v_{\theta}}{v_r} = \frac{k_r}{k_{\theta}},
 \end{equation}
where $k_r$ and $k_{\theta}$ are the IGW radial and tangential wavenumbers, respectively. Clearly, this expression does not prove the presence of non-linear waves, so satisfying this relationship indicates the presence of only linear waves. Taking the log of both sides of Eq.~\ref{eq:ratio_vr_vtheta}, replacing $k_r$ and $k_{\theta}$ with the appropriate expressions for IGW wavenumbers, and differentiating both sides in terms of $\partial\log_{10}\omega$, we obtain 
 \begin{equation}\label{eq:tan_rad_vel_relation}
     \frac{\partial \log_{10}v_{\theta}}{\partial \log_{10} \omega} = \frac{\partial \log_{10}v_{r}}{\partial \log_{10} \omega} - 1. 
 \end{equation}
The above expression indicates that for linear waves, the slopes of the radial velocity and tangential velocity should differ by 1, and this is what we observe approximately in Fig.~\ref{fig:vtheta-vr-tem-grid513}. For example, the $v_r$ slope for 3\msol{} is -0.94 and that for $v_{\theta}$ is -2.06. It is obvious that we should not expect the difference to be perfectly equal to 1, which is the case for all the stellar models, as we are fitting stochastic perturbations. However, the fact that the difference is close to 1 is reassuring.  

\begin{table*}[ht!]
\caption{Frequency slopes of different stellar quantities close to the top of the simulation domain. The error represents changes due to our selection of straight-line fitting frequency range and the different tangential direction at which the quantity is measured. }
\label{tab:slopes_max_min_1400}
\centering
        \begin{tabular}{cccc}
            \hline\hline
            Model & Radial Velocity & Tangential Velocity & Temperature  \\
            \hline
            3 \msol{} midMs &  -0.088 $\pm$ 0.455 & -1.197 $\pm$ 0.470 & -1.099 $\pm$ 0.456 \\
            5 \msol{} midMs &  -0.229 $\pm$ 0.634 & -1.293 $\pm$ 0.347 & -1.216 $\pm$ 0.622 \\
            7 \msol{} midMs &  -0.864 $\pm$ 0.476 & -1.783 $\pm$ 0.503 & -1.856 $\pm$ 0.475 \\
            10 \msol{} midMs &  -0.174 $\pm$ 0.379 & -1.384 $\pm$ 0.350 & -1.178 $\pm$ 0.344 \\
            13 \msol{} midMs &  -0.619 $\pm$ 0.979 & -1.574 $\pm$ 0.495 & -1.529 $\pm$ 0.896 \\
            \hline
        \end{tabular}%}
\end{table*}

\begin{table*}[ht!]
\caption{Frequency slopes of different stellar quantities at 0.25 $H_P$ from the boundary where the \bvf{} goes to 0. The error represents changes due to our selection of straight-line fitting frequency range and the different tangential direction at which the quantity is measured.}
\label{tab:slopes_max_min_513}
\centering
        \begin{tabular}{cccc}
            \hline\hline
            Model & Radial Velocity & Tangential Velocity & Temperature  \\
            \hline
            3 \msol{} midMs &  -0.248 $\pm$ 0.928 & -1.290 $\pm$ 0.984 & -1.297 $\pm$ 0.838 \\
            5 \msol{} midMs &  -0.438 $\pm$ 0.801 & -1.525 $\pm$ 0.842 & -1.422 $\pm$ 0.728 \\
            7 \msol{} midMs &  -0.897 $\pm$ 0.766 & -1.955 $\pm$ 0.712 & -1.832 $\pm$ 0.782 \\
            10 \msol{} midMs &  -0.448 $\pm$ 0.578 & -1.592 $\pm$ 0.662 & -1.407 $\pm$ 0.583 \\
            13 \msol{} midMs &  -0.551 $\pm$ 0.809 & -1.388 $\pm$ 0.941 & -1.465 $\pm$ 0.795 \\
            \hline
        \end{tabular}%}
\end{table*}

Returning to Fig.~\ref{fig:ratio-vtheta-vr}, the bottom two panels show the ratio, $v_{\theta}/v_r$, above the middle of the radiation zone and close to the top of the simulation domain. We can see that except at very low frequencies, the linear wave properties are still observed\footnote{There is a minimal shift of the ratio for all the models above the solid, black line in the last row, mainly because close to the simulation boundary, the radial velocity is being artificially forced to 0. Thus, the radial velocity generally starts to drop, resulting in the increase in the ratio of velocities.}. These waves should, then, propagate through the radiation zone and manifest themselves close to the top of the domain, with properties that can be explained with linear wave theory. In Fig.~\ref{fig:vtheta-vr-tem-grid1400}, we show the velocity and temperature spectra close to the top of the simulation domain, where we see a general trend of smaller variations and lower amplitudes at lower frequencies. As the frequency increases, we observe larger variations and higher amplitudes at mid-range frequencies, but with a negative slope for all the stellar models. 

The spectra that we have presented so far are related to a radial ray, chosen at a random horizontal angle. Due to the stochastic behaviour of these perturbations, variations are expected for radial rays at different angles. Additionally, we chose a fixed frequency range (between 8 \micromu{} and 30 \micromu{}) to calculate the slopes for all the models, so we expect further variations for different frequency intervals. Thus, in Table~\ref{tab:slopes_max_min_1400}, we present the average slopes with standard deviations for the radial velocities, tangential velocities, and temperatures close to the surface of each model (see the panels at the bottom of Fig.~\ref{fig:ratio-vtheta-vr} for the exact radii). These slope averages and standard deviations are calculated for radial rays at different angles and different frequency intervals, all starting at 8 \micromu{}.  Equation~\ref{eq:tan_rad_vel_relation} is approximately fulfilled for the slopes of the radial and tangential velocities. Finally, as seen in the figures shown in this section, the temperature slopes are still similar to the tangential velocity slopes. Comparing these slopes with the slopes at the convective-radiative interface (as shown in Fig.~\ref{fig:vtheta-vr-tem-grid513}), s similar behaviour is observed for the averaged slopes, as shown in Table~\ref{tab:slopes_max_min_513}. This table  suggests that during propagation in the stellar interior, IGWs do not experience any frequency dependent changes within the chosen frequency range here. However, the amplitude of the perturbations in Figs.~\ref{fig:vtheta-vr-tem-grid513} and \ref{fig:vtheta-vr-tem-grid1400} do show that there is a constant increase at all frequencies as IGWs propagate, which is consistent with the predictions of linear IGW theory (see $A_{0}$ in Eq.~\ref{eq:A_exp_tau}, where $A_{0}$ can be dependent on stellar radius).

\section{Comparison with photometric data}\label{sec:Compare}
\begin{table*}
\caption{Best frequency slope matches for the different TESS models studies with the temperature perturbations from our simulations. For more details on the periodograms of the TESS models, we refer to Appendix~A.}
\label{tab:slopes_compare}
\centering
        %\resizebox{\textwidth}{!}{%
        \begin{tabular}{cccc}
            \hline\hline
            Simulation model & TESS star mass/\msol{} & TESS star TIC & Frequency slope \\%& Frequency Range/\micromu{}\\
            \hline
            3\msol{} midMs & 2.51 & 032664824  & -0.86 \\%& 16.4 $\leqslant f \leqslant$ 40.9\\
            & 2.68 & 175173003 & -1.35 \\%& 17.3 $\leqslant f \leqslant$ 43.2 \\
            \hline
            5\msol{} midMs & 4.02 & 178379696 & -0.98 \\%&  15.5 $\leqslant f \leqslant$ 38.7\\
            & 4.58 & 028949811 & -0.90 \\%&  23.3 $\leqslant f \leqslant$ 58.3\\
            & 5.50 & 048541963 & -1.03 \\%&  22.8 $\leqslant f \leqslant$ 57.0\\
            & 5.99 & 260646994 & -0.74 \\%&  8.7 $\leqslant f \leqslant$ 21.6\\
            & 5.99 & 412543692 & -1.27 \\%&  22.6 $\leqslant f \leqslant$ 56.5\\
            \hline
            7\msol{} midMs & 6.18 & 439430953& -1.48 \\%&  15.0 $\leqslant f \leqslant$ 47.1 \\
            & 7.55 & 132916101 & -1.87 \\%&  10.7 $\leqslant f \leqslant$ 33.5 \\
            %& 8.52 & 393159663 & -1.41 \\%&  16.7 $\leqslant f \leqslant$ 52.5 \\
            \hline
                10\msol{} midMs & 9.55 & 380435021 & -0.90 \\%&  12.6 $\leqslant f \leqslant$ 39.5\\
            & 9.82 & 438967685 & -1.00 \\%&  11.0 $\leqslant f \leqslant$ 34.5\\
            & 10.10 & 199833554 & -1.09 \\%&  10.8 $\leqslant f \leqslant$ 33.8 \\
            & 10.47 & 315506577 & -1.18 \\%&  8.7 $\leqslant f \leqslant$ 27.2\\
            & 10.50 & 205871617 & -1.21 \\%&  9.1 $\leqslant f \leqslant$ 28.5 \\
            & 10.79 & 461021821 & -1.29 \\%& 12.6 $\leqslant f \leqslant$ 31.4 \\
                & 10.96 & 466592679 & -0.96 \\%& 16.2 $\leqslant f \leqslant$  40.4 \\ 
            \hline
            13\msol{} midMs & 11.26 & 078636551 & -0.90 \\%& 10.2 $\leqslant f \leqslant$  25.4 \\
                & 11.31 & 448888783 & -1.22 \\%&  9.4 $\leqslant f \leqslant$ 23.4\\
                & 11.80 & 322418931 & -1.12 \\%&  9.8 $\leqslant f \leqslant$ 24.4\\
                & 12.60 & 140129494 & -1.41 \\%&  8.0 $\leqslant f \leqslant$ 19.8\\
                & 12.96 & 011400562 & -1.47 \\%& 8.5 $\leqslant f \leqslant$ 26.6 \\
                & 13.20 & 065516748 & -1.40 \\%& 8.9 $\leqslant f \leqslant$ 27.8\\
            \hline
        \end{tabular}%}
\end{table*}
The velocity and temperature perturbations seen at the top of our simulation domain are likely to be related to brightness variations seen in photometric data from stars. One of the first direct comparisons for intermediate-mass stars was done in \cite{2015ApJ...806L..33A}. This work noted the similarity between the velocity perturbations from the simulations \citep{2013ApJ...772...21R} of a 3 \msol{} star, with the brightness variations from young massive O-type stars observed with high-precision CoRoT space photometry. Many works from both the numerical and observation fronts followed, but a direct comparison between a set of refined photometric data and simulation results of different masses remains unexplored. To address this, we utilised some results of a recent survey of $\beta$ Cephei stars in the Southern Hemisphere, conducted by a few co-authors of this paper, and whose results are yet to be published. In the survey, photometric data from the TESS satellite were used along with spectroscopy to classify and carry out statistical studies of $\beta$ Cephei stars. In total, 486 stars were observed in low resolution (421 stars) and high resolution (65 stars) spectroscopic mode, and their spectroscopic parameters were derived, to complement the photometric aspect of the survey. Both photometrically and spectroscopically, the target selection was made through various means, which will be described in detail in Handler et al.\ (in prep) and Chowdhury et al.\ (in prep). To derive luminosities for these 486 spectroscopically observed stars, GAIA EDR3 parallaxes and TIC V magnitudes were utilized. All bolometric corrections were applied using interpolated values derived from \citet{Flower1996}. The extinction was calculated based on \textsc{mwdust} python package \citep{Bovy2016}, using the combined \citep{Marshall2006, Green2015, Drimmel2003} reddening map provided by \cite{Bovy2016}. 

Having their T$_{\mathrm{eff}}$ and luminosity values derived, these 486 stars were then plotted in the HR diagram with the evolutionary tracks calculated using the Warsaw-New Jersey stellar evolution code (for details see, e.g. \cite{Pamyatnykh1999}, and references therein). To calculate the tracks, they used OPAL opacity tables, hydrogen abundance (X) as 0.74, heavy element abundance (Z) as 0.02, and initial rotational velocity as 150 km s$^{-1}$ (sample average). The masses were then extracted by comparing the position of stars in the HR diagram, with the evolutionary tracks for 61 different masses between 2.5 \msol{} and 60 \msol{} with varying step size, and nine different rotational velocities between 0 and 350 km s$^{-1}$ with a step size of 50 km s$^{-1}$. For this work, we utilize these derived masses.

Before comparing our simulation data with the observational data, we have to consider a few issues in choosing the sample. Our output data is 2D in space: radial and tangential. Thus, for a time-series analysis at any particular radius, there is freedom with respect to choosing the angle at which this is done. For the purposes of the analysis of a single radial ray (inside the convection zone, close to the interface, and at the top of the simulation domain), observations at a few angles and obtaining an error estimate would suffice (as shown in the previous section). However, for the purpose of comparing our simulation results with the photometric data of stars, a few approaches can be considered, but we chose to perform a weighted average of spectra over half the domain circumference. This is done to account for the fact that telescopes receive light from only half of any star that is observed. Additionally, we receive fewer photons from the edges of a star compared to the centre. To account for all of these effects, we performed a weighted average as shown in the equation below:
\begin{equation}
    f_{avg} = \frac{1}{N}\sum^{N}_{i} f_i - f_ib(1-\sin \theta),  
\end{equation}
where $N$ is the total number of angular grid points over half the domain circumference, $b$ is a decimal number between 0 and 1, and $\theta$ is the angle of observation from the edge of half the simulation domain. We chose a simple linear expression to represent the photon count drop due to the geometrical effect mentioned above and a value of 0.6 for the coefficient $b$.  

As mentioned earlier, the data from the survey done using the TESS satellite data covers a variety of stars at different masses. Thus, we compared our simulation results with the observational data in the following manner: 2 \msol{} - 4 \msol{} with 3 \msol{} midMs, 4 \msol{} - 6 \msol{} with 5 \msol{} midMs, 6 \msol{} - 9 \msol{} with 7 \msol{} midMs, 9 \msol{} - 11 \msol{} with 10 \msol{} midMs, and 11 \msol{} - 15 \msol{} with 13 \msol{} midMs. The next step in the process was to decide which of the three simulation stellar parameters was best to compare with the brightness variations seen in the TESS photometric data. For this work, we make a choice of comparing the temperature perturbations, in units of Kelvin, with the brightness variations, in units of \si{\micro mag}, seen in the TESS photometric data. The temperature perturbations were chosen mainly because of direct relations between stellar luminosity and temperature (i.e. Stefan-Boltzmann law).

The main objective of this part of the study was to search for similarities and differences between the slopes of the brightness variations from the TESS photometric data and the slopes of temperature perturbations in our simulation results. To achieve this, we selected a fixed frequency window in logarithmic space and calculated the slopes of both the temperature perturbations and brightness variations with straight line fits within this frequency range. We then moved the window along the frequency axis in small increments and repeated the slope calculation. This scanning process was done from 8 \micromu{} to 100 \micromu{}, for different frequency intervals. The frequency range at which the difference between the temperature perturbation slope and the brightness variation slope is less than 5\% (chosen arbitrarily) was selected as the best match. Due to our strict tolerance value, the number of stars that we found suitable for comparison reduced from the initial 486 in the survey to 22, with more higher mass stars fitting our criteria than lower mass stars. 

The periodograms of the TESS data are over-plotted on the temperature perturbations from the simulations and presented in Appendix~A. An important point to note here is that due to the difference in the units of the brightness variations and temperature perturbations, we adjusted the temperature perturbation amplitudes by using the y-intercepts of the straight-line fits done to these data. We also present the best matches for frequency slopes, the frequency range for these slopes, and the stellar masses for the simulation models and TESS stars in Table~\ref{tab:slopes_compare}. The most negative frequency slope is found to be -1.87 and the least negative frequency slope is -0.74, with most of the slopes being more negative than -1.0. We found that for most of the TESS data, the frequency range at which the simulated temperatures agree best with the TESS surface brightness variations, lies above 10 \micromu{}. For higher mass stars, however, this frequency range shifts to lower values ($\geqslant$ 8 \micromu{}), which, with the assumption that the TESS brightness variations are due to IGWs from core convection, could be caused by the decrease in the average \bvf{} frequency when the stellar mass is increased. In some cases (e.g. model TIC 412543692, 461021821, 466592679, 078636551 and 065516748), the periodograms even show that the amplitudes start to rise\footnote{These indicate p-modes, which are the unique characteristic of $\beta$-Cepheid stars. The anelastic simulations do not allow propagation of sound waves, thus inhibiting p-mode formation.}(orange lines), when simulated temperatures (blue lines) start to fall rapidly due to the \bvf{} frequency limit. 

\section{Discussion \& conclusions}
We ran 2D simulations to study the internal fluid evolution of five stellar models: 3\msol{}, 5\msol{}, 7\msol{}, 10\msol{}, and 13 \msol{}, which are all chosen at the middle of  the main sequence (midMs). These stars all possess a convective core and a radiative envelope, which is cut off at 90\% of the total stellar radius, to ensure numerical stability. Compared to previous studies with such anelastic hydrodynamical simulations, where constant thermal diffusivities (as a function of radius) were implemented, this study utilises varying thermal diffusivity profiles that are more realistic and allows for  any additional effect due to the variations in thermal diffusivity to exist.  

Before analysing our simulation results, we used a simple linear theory model for internal gravity waves (IGWs) to calculate the effect of thermal diffusivity profiles that we chose for our models. We found that in the case of linear IGWs, any perturbations above 8 \micromu{}, will be approximately represented up to 50\% accuracy, compared to the same perturbations with the real stellar thermal diffusivities. Thus, we chose 8 \micromu{} as the lower limit for all of our analysis and we proceeded to study the radial velocity, tangential velocity, and temperature perturbations in frequency space. One of our main conclusions is that  the perturbations generally show linear IGW-like behaviour throughout the radiation zone. Second, the frequency slopes at the domain surface are, on average, $< -1$ for the tangential velocity and temperature perturbations and $> -1$ for the radial velocity perturbation. This is also true at the convective-radiative interface. Third, there is no clear trend for these frequency slopes across the different stellar masses, although there is a clear trend of increasing amplitudes with increasing mass, mainly due to the higher convective velocities for higher stellar masses. 

Although we did not delve extensively into the topic of IGW excitation in this work, we would like to offer a remark on this here. Generally, we observe that plumes are dominant in the convection zone (similar to \cite{2013ApJ...772...21R}). This is mainly seen from the IGW frequency slopes at generation, which are all shallower than -2 (between 0 and -2). These flatter slopes agree well with the theoretical work on plume forcing done in \cite{pincon2016a}, and the numerical work done in \cite{philipp3dpaper} and \cite{Horst2020} on IGW generation. Earlier works such as that of \cite{1999ApJ...520..859K}, where the authors carried out a theoretical study of  IGWs generated through Reynolds' stresses inside the convection zone and then tunnel through predicted slopes steeper than -3, whilst \cite{2013MNRAS.430.2363L} predicted slopes steeper than -4. Both these works on IGW generation through Reynolds' stress predict much steeper slopes than what we have observed, which leads us to make the claim that the IGWs in our simulations are generated primarily through plumes.

In the previous section, we attempted to utilise the photometric data from a recent survey  by several co-authors of this paper to directly compare the frequency slopes from these observations with the frequency slopes from our simulation. We chose the slopes from the simulated temperature perturbations and found the best slope match for multiple radial IGW rays (which vary with horizontal angles), and frequency intervals. For these 26 stars in particular, which are mainly a mix of $\beta$-Cepheid variables, slowly pulsating B-type stars, and stars which just typically show low-frequency variability, the best-match frequency slopes range from -1.87 to -0.74, which we found for different frequency intervals, all starting from 8 \micromu{}. 

This work mainly shows the low-frequency variabilities, seen in the photometric data of some stars, could be explained by the IGWs from core convection. Our results indicate that the surface frequency slopes are consistent with the observational results to-date \citep{dom_igw_exp_range,dom2020}. We found a more recent observational work, \cite{dom_igw_exp_range} -- where it was found that the frequency slope magnitudes of IGW signatures from a variety of blue supergiants (BSGs) are less than
3.5 -- to be the most comparable to the findings of the present work. Unfortunately, we were not able to extract information about the stellar mass from our simulated data. The largest differences between the models that we used lie in the stellar radius, density, \bvf{} frequency, and thermal diffusivity profiles (which introduces 4 degrees of freedom, calling for a study on how to reduce this). In addition, investigating the effect of varying just one parameter while keeping the others constant would be beneficial, albeit tedious. 

In this work, we used a constant solid-body rotation rate for all the simulations. However, O- and B-type stars are known to have angular velocities that vary with radius and polar angles. Thus, the next natural step for this work would be to study the effect of differential rotation for different masses. Studies on the propagation of IGWs through a radial differential rotation have already been done on one stellar mass \citep{2013ApJ...772...21R}, but repeating this for a broader range of masses is crucial as we do not expect the rotation profile to be the same for all stellar masses. This study on differential rotation will also greatly help us understand the transport of angular momentum and chemical composition, which are two of the more poorly understood fundamental problems in stellar evolution.

\begin{acknowledgements}
       PVFE was supported by the U.S. Department of Energy through the Los Alamos National Laboratory (LANL). LANL is operated by Triad National Security, LLC, for the National Nuclear Security Administration of the U.S. Department of Energy (Contract No. 89233218CNA000001). This work has been assigned a document release number LA-UR-22-26062.
\end{acknowledgements}

% WARNING
%-------------------------------------------------------------------
% Please note that we have included the references to the file aa.dem in
% order to compile it, but we ask you to:
%
% - use BibTeX with the regular commands:
%   \bibliographystyle{aa} % style aa.bst
%   \bibliography{Yourfile} % your references Yourfile.bib
%
% - join the .bib files when you upload your source files
%-------------------------------------------------------------------
\bibliographystyle{aa}
\bibliography{bi}

\begin{appendix} %First appendix
\section{Periodograms of TESS stars versus simulated temperature perturbations}
We present the simulated surface temperature perturbations from multiple stellar models, 3 \msol{} to 13 \msol{} (see Table~\ref{tab:slopes_compare}) in blue, overplotted on the periodograms of selected stars, and in orange, from the TESS catalogue, specifically: from the study in Handler et al.\ (in prep) and Chowdhury et al.\ (in prep). The titles of each plot provide the TIC number and mass calculated from HR diagram fitting (see Section~\ref{sec:Compare}). The legends show the mass of the stars used in our simulations. The y-axes represent brightness perturbations, where we have converted the temperature perturbations from simulations using the multiplication factor shown in brackets in the legend to align both datasets.
\begin{figure}[ht!]
        \centering
        \includegraphics[trim={0.0cm 0.0cm 0 0.0cm},clip,width=0.9\columnwidth]{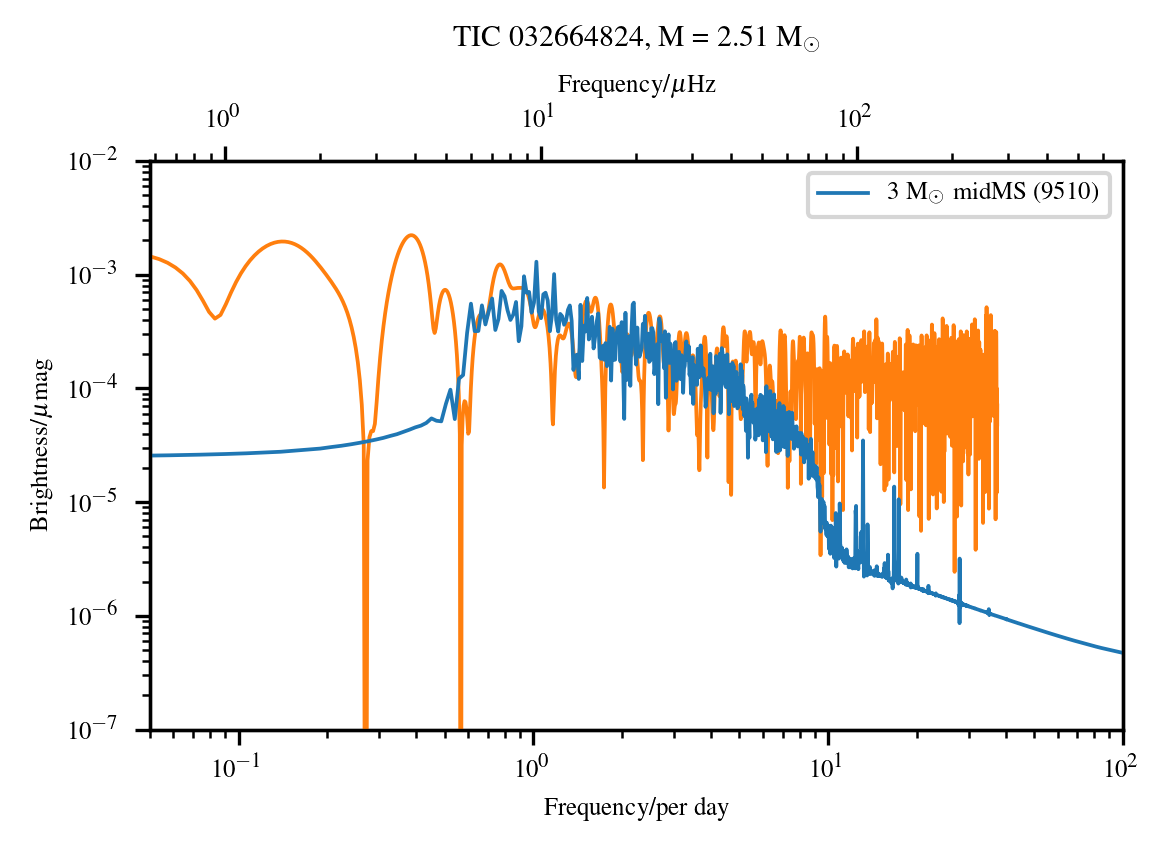}
        \caption{Normalised temperature perturbation (solid blue line) from a 3 \msol{} star simulation and brightness variation (orange, solid line) from a 2.51 \msol{} star from the TESS catalogue as a function of wave frequencies.}
\end{figure}
\begin{figure}[ht!]
        \centering
        \includegraphics[trim={0.0cm 0.0cm 0 0.0cm},clip,width=0.9\columnwidth]{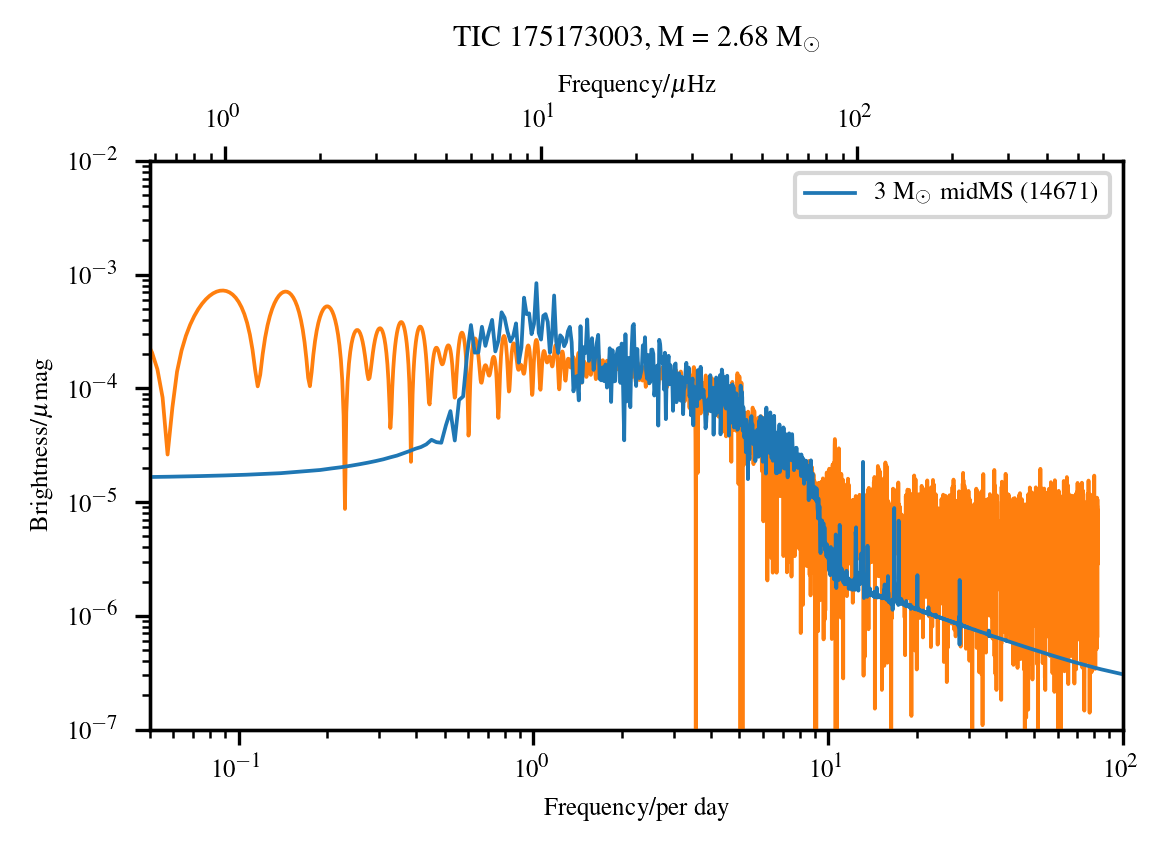}
         \caption{Normalised temperature perturbation (solid blue line) from a 3 \msol{} star simulation and brightness variation (orange, solid line) from a 2.68 \msol{} star from the TESS catalogue as a function of wave frequencies.}
\end{figure}

\begin{figure}[ht!]
        \centering
        \includegraphics[trim={0.0cm 0.0cm 0 0.0cm},clip,width=0.9\columnwidth]{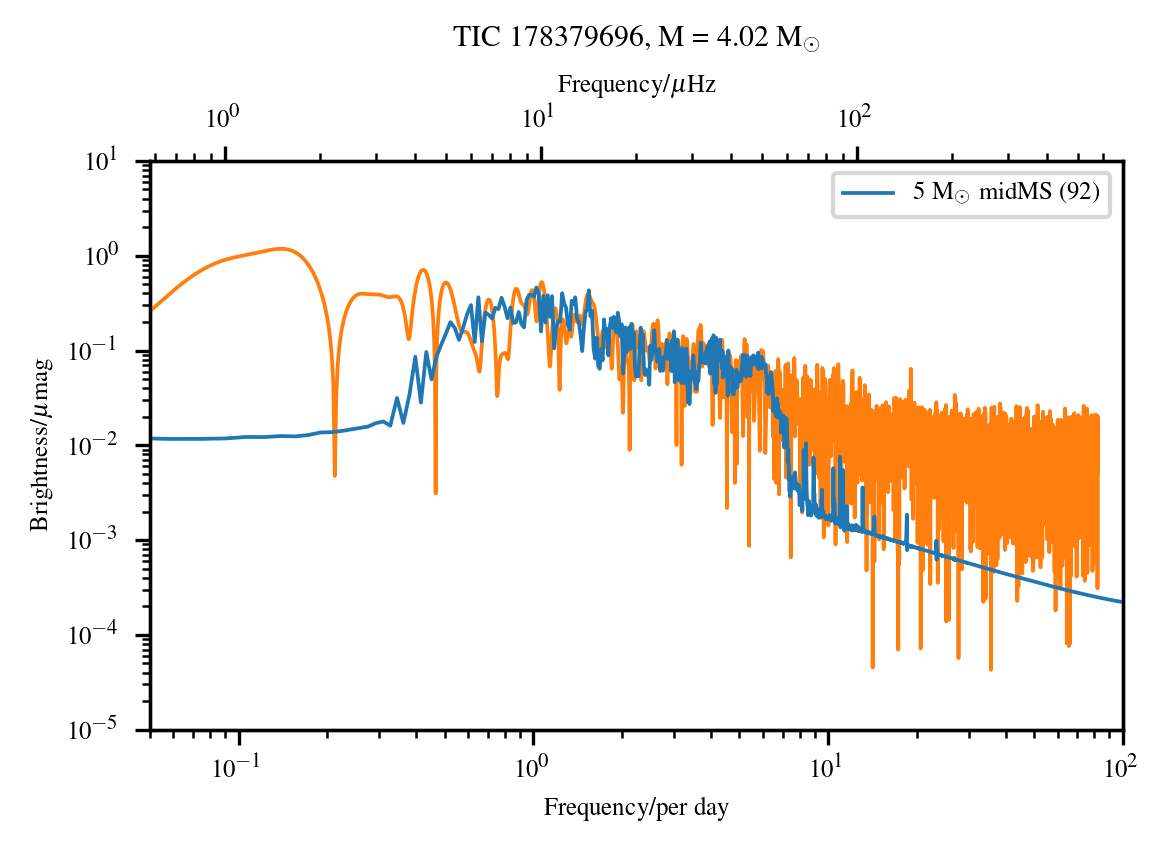}
        \caption{Normalised temperature perturbation (solid blue line) from a 5 \msol{} star simulation and brightness variation (orange, solid line) from a 4.02 \msol{} star from the TESS catalogue as a function of wave frequencies.}
\end{figure}
\begin{figure}[ht!]
        \centering
        \includegraphics[trim={0.0cm 0.0cm 0 0.0cm},clip,width=0.9\columnwidth]{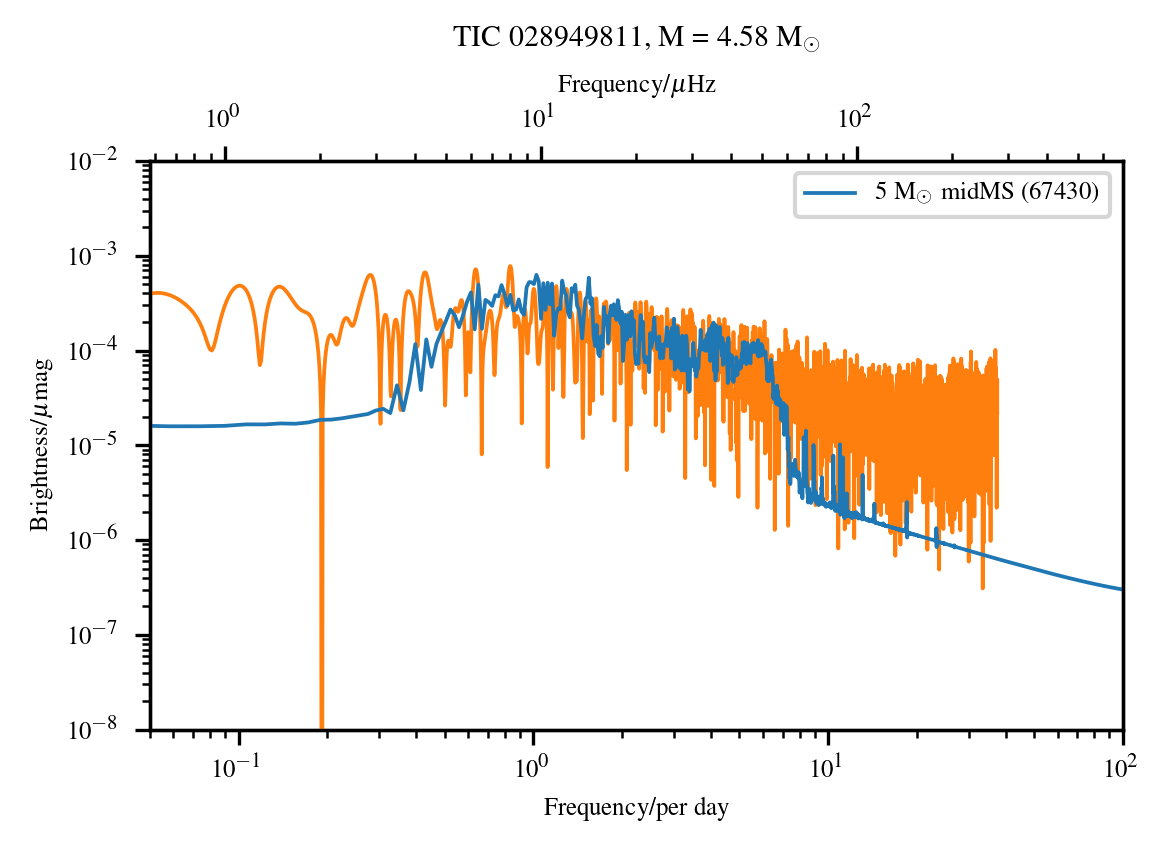}
        \caption{Normalised temperature perturbation (solid blue line) from a 5 \msol{} star simulation and brightness variation (orange, solid line) from a 4.58 \msol{} star from the TESS catalogue as a function of wave frequencies.}
\end{figure}
\begin{figure}[ht!]
        \centering
        \includegraphics[trim={0.0cm 0.0cm 0 0.0cm},clip,width=0.9\columnwidth]{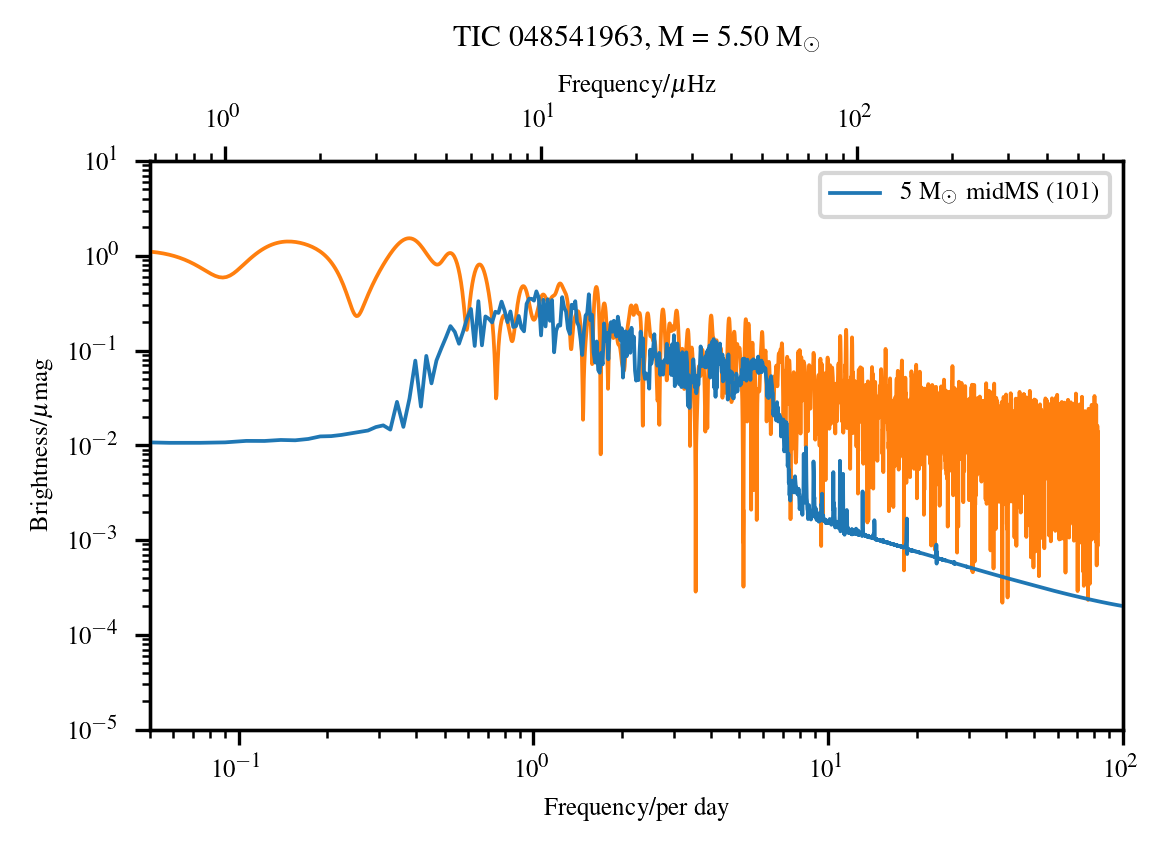}
         \caption{Normalised temperature perturbation (solid blue line) from a 5 \msol{} star simulation and brightness variation (solid orange
line) from a 5.50 \msol{} star from the TESS catalogue as a function of wave frequencies.}
\end{figure}
\begin{figure}[ht!]
        \centering
        \includegraphics[trim={0.0cm 0.0cm 0 0.0cm},clip,width=0.9\columnwidth]{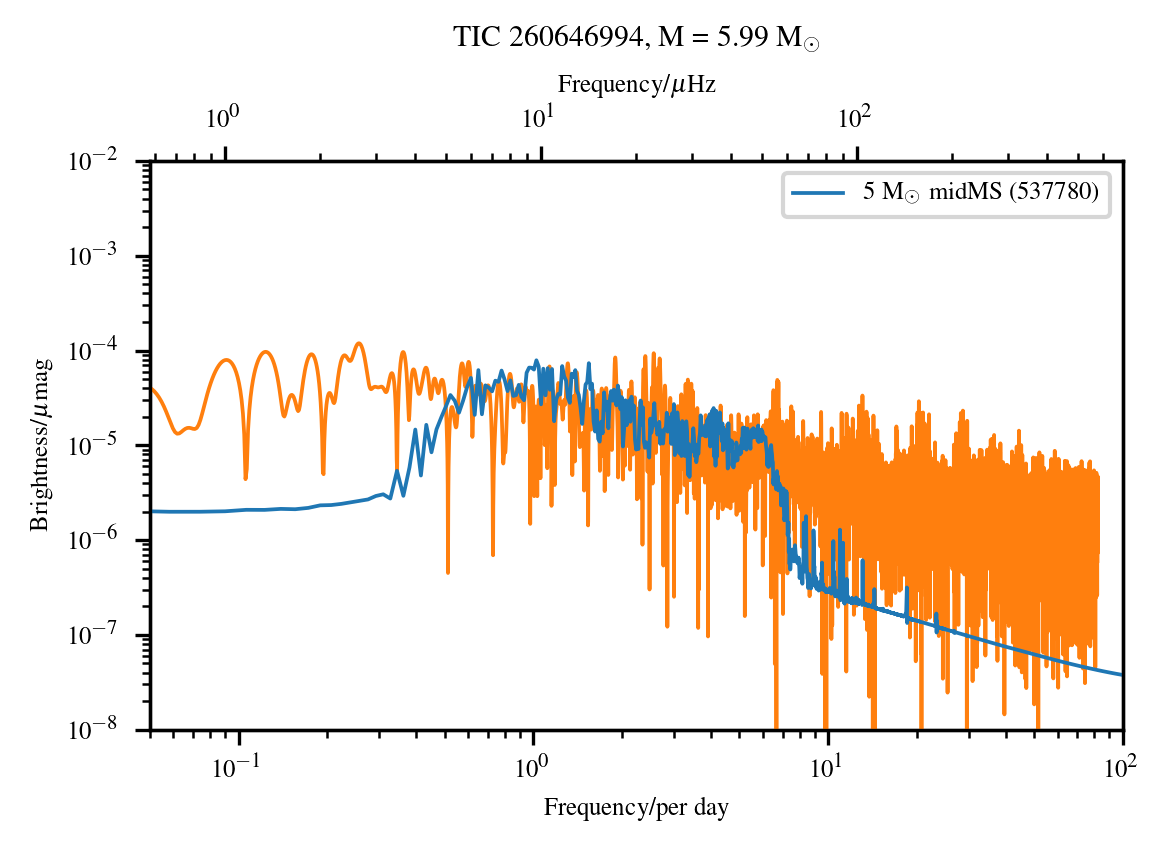}
         \caption{Normalised temperature perturbation (solid blue line) from a 5 \msol{} star simulation and brightness variation (solid orange
line) from a 5.99 \msol{} star from the TESS catalogue as a function of wave frequencies.}
\end{figure}
\begin{figure}[ht!]
        \centering
        \includegraphics[trim={0.0cm 0.0cm 0 0.0cm},clip,width=0.9\columnwidth]{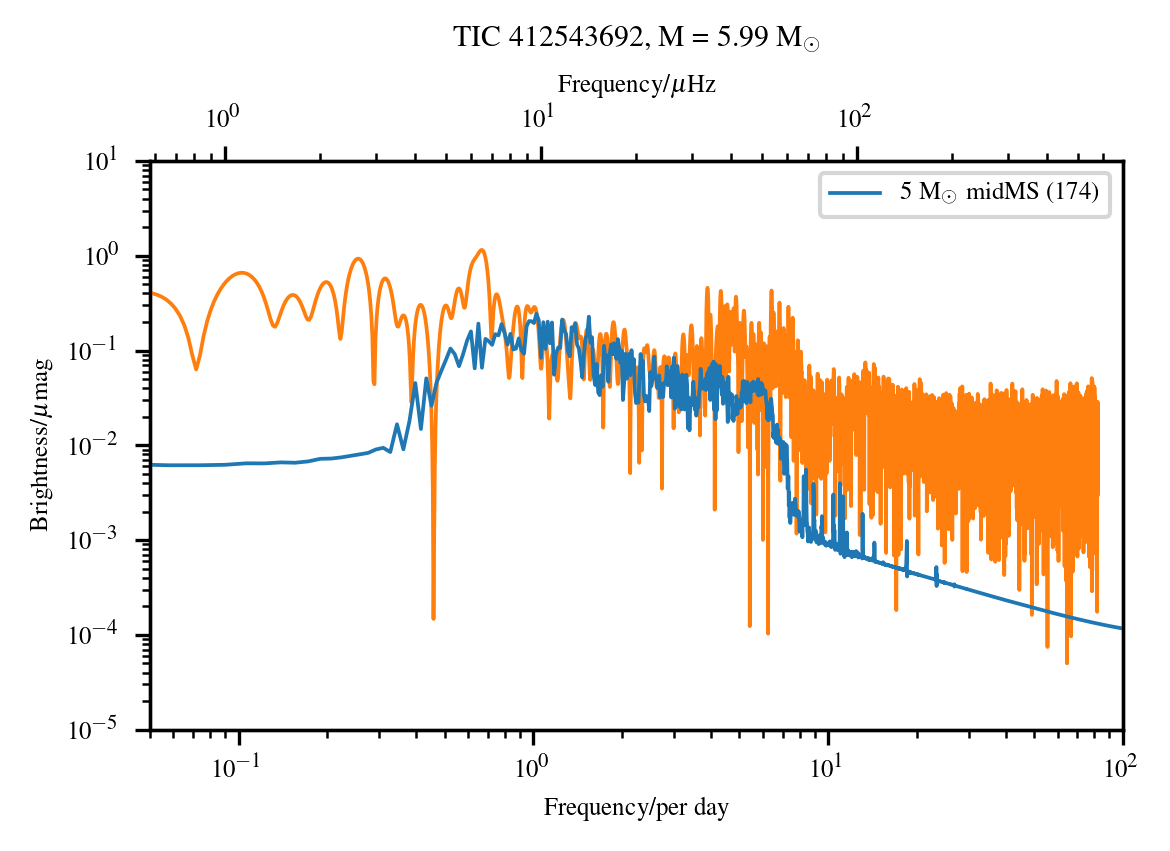}
         \caption{Normalised temperature perturbation (solid blue line) from a 5 \msol{} star simulation and brightness variation (solid orange
line) from a 5.99 \msol{} star from the TESS catalogue as a function of wave frequencies.} 
\end{figure}

\begin{figure}[ht!]
        \centering
        \includegraphics[trim={0.0cm 0.0cm 0 0.0cm},clip,width=0.9\columnwidth]{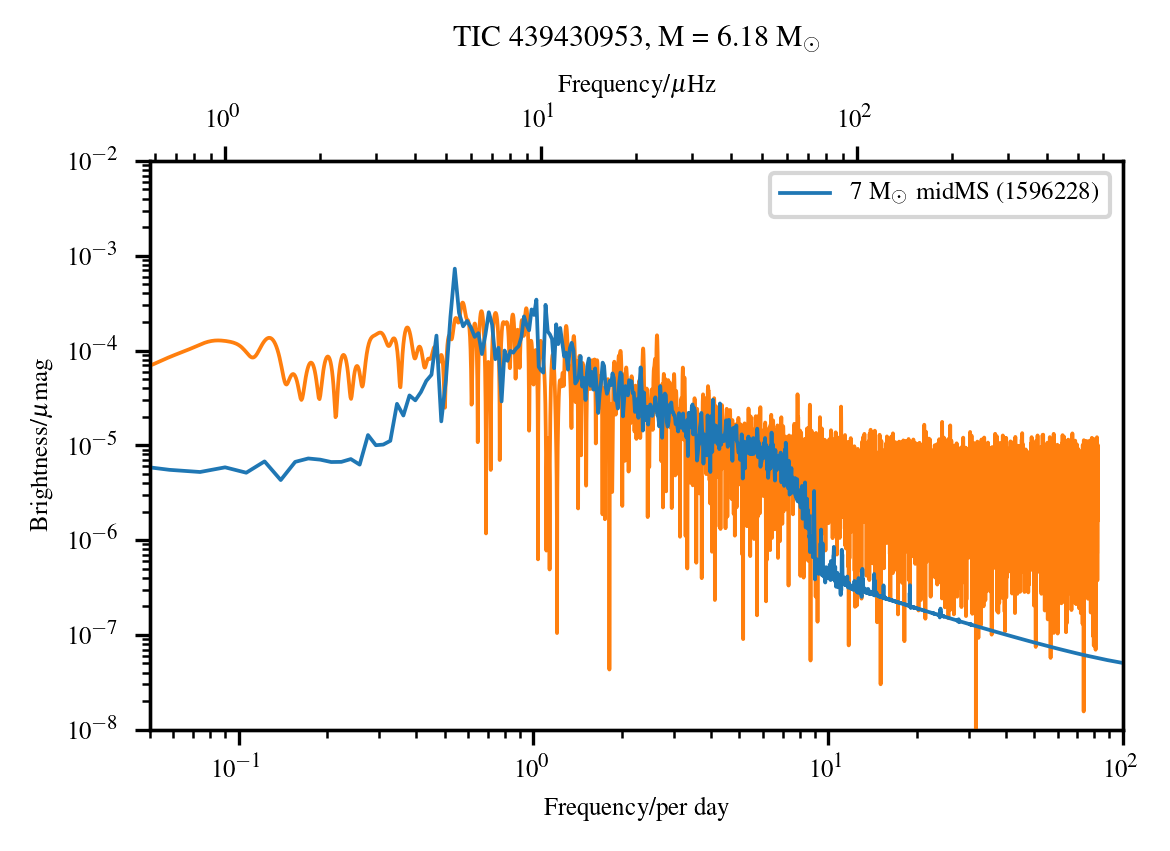}
         \caption{Normalised temperature perturbation (solid blue line) from a 7 \msol{} star simulation and brightness variation (solid orange
line) from a 6.18 \msol{} star from the TESS catalogue as a function of wave frequencies.}
\end{figure}
\begin{figure}[ht!]
        \centering
        \includegraphics[trim={0.0cm 0.0cm 0 0.0cm},clip,width=0.9\columnwidth]{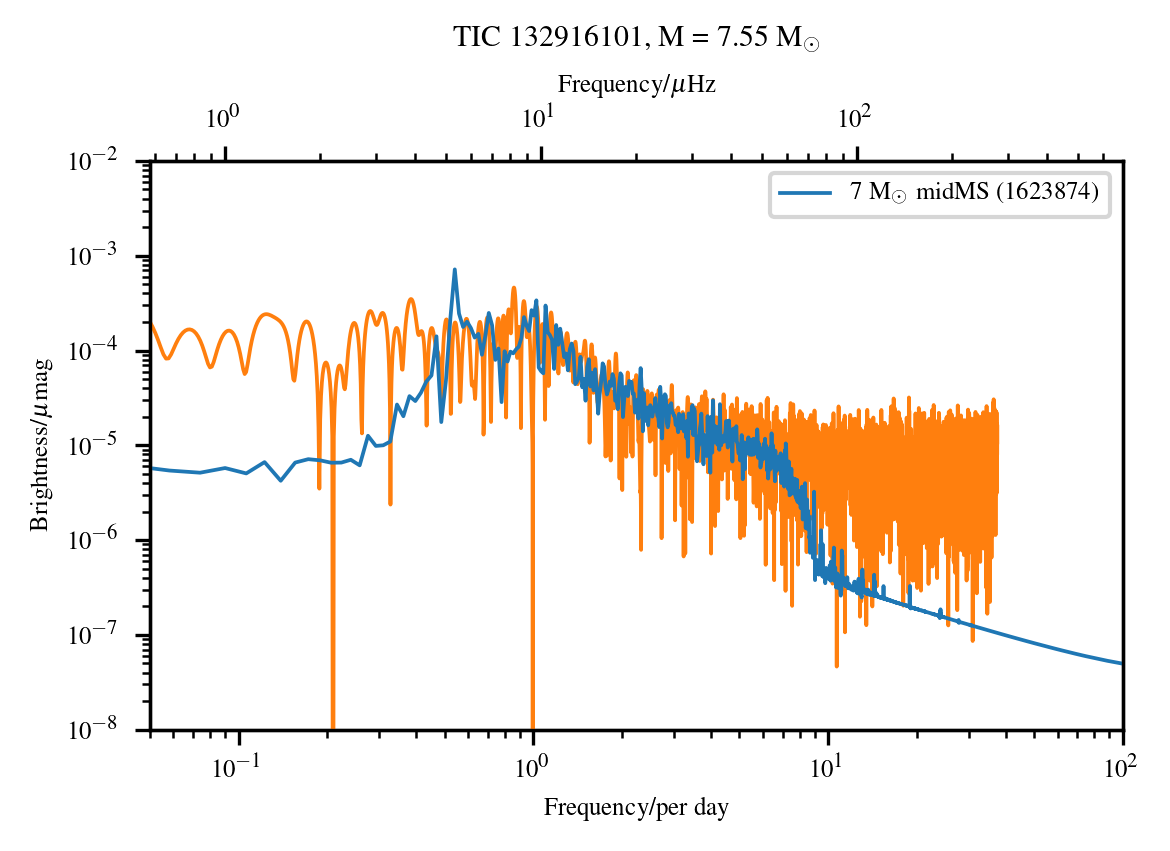}
          \caption{Normalised temperature perturbation (solid blue line) from a 7 \msol{} star simulation and brightness variation (solid orange
line) from a 7.55 \msol{} star from the TESS catalogue as a function of wave frequencies.}
\end{figure}
\begin{figure}[ht!]
        \centering
        \includegraphics[trim={0.0cm 0.0cm 0 0.0cm},clip,width=0.9\columnwidth]{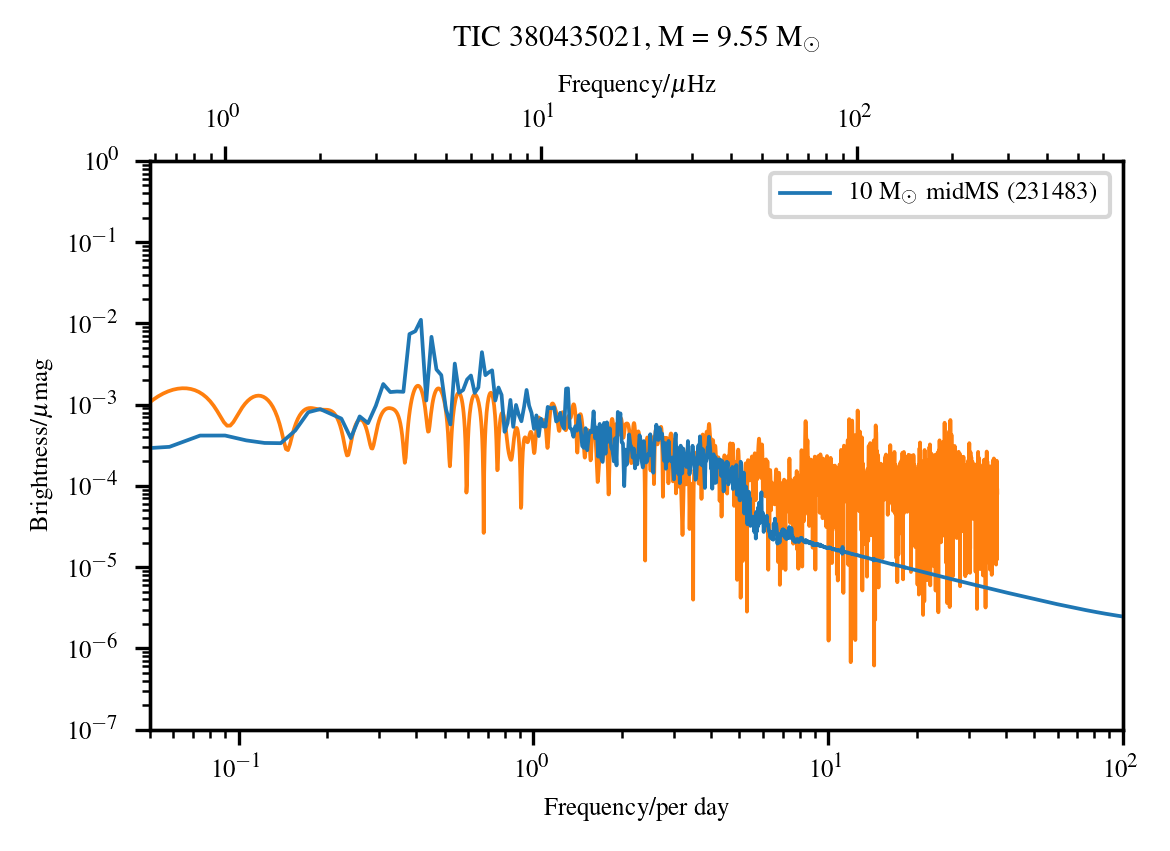}
          \caption{Normalised temperature perturbation (solid blue line) from a 10 \msol{} star simulation and brightness variation (solid orange line) from a 9.55 \msol{} star from the TESS catalogue as a function of wave frequencies.}
\end{figure}
\begin{figure}[ht!]
        \centering
        \includegraphics[trim={0.0cm 0.0cm 0 0.0cm},clip,width=0.9\columnwidth]{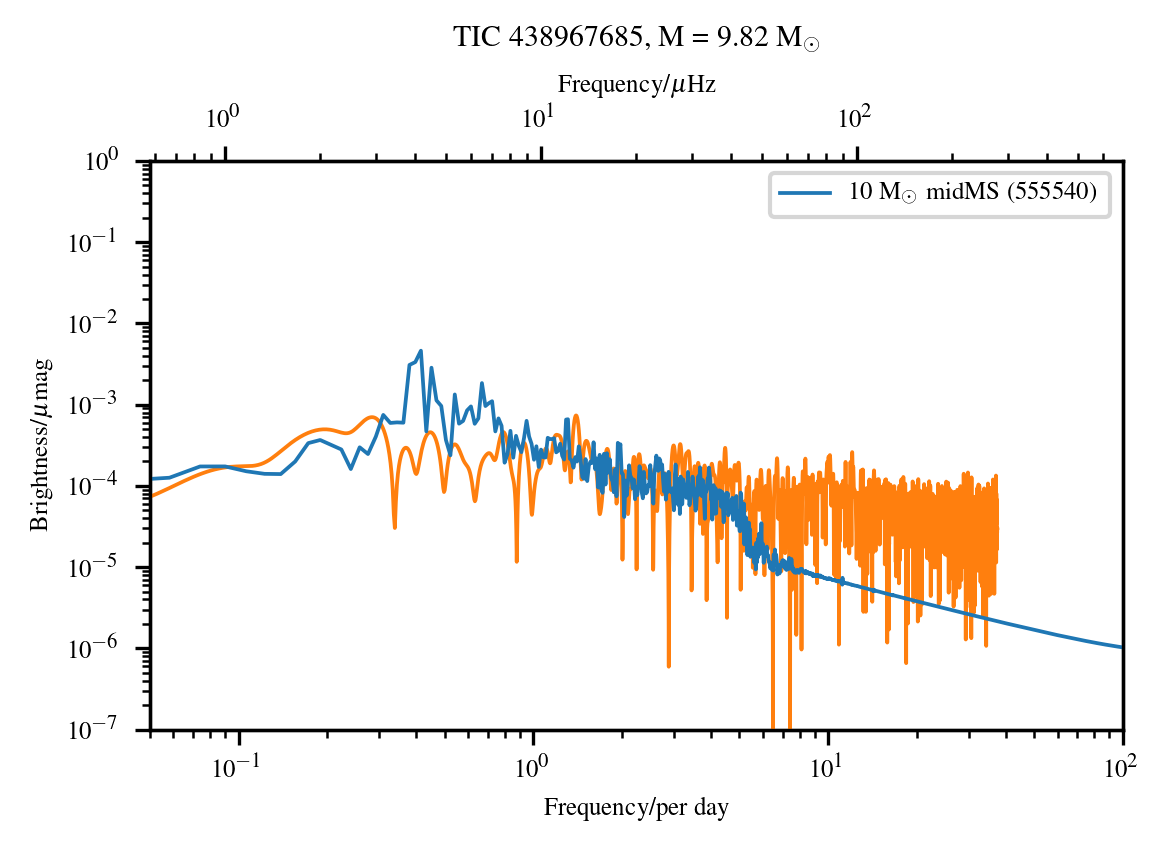}
        \caption{The normalised temperature perturbation (solid blue line) from a 10 \msol{} star simulation and brightness variation (solid orange
line) from a 9.82 \msol{} star from the TESS catalogue as a function of wave frequencies.}
\end{figure}
\begin{figure}[ht!]
        \centering
        \includegraphics[trim={0.0cm 0.0cm 0 0.0cm},clip,width=0.9\columnwidth]{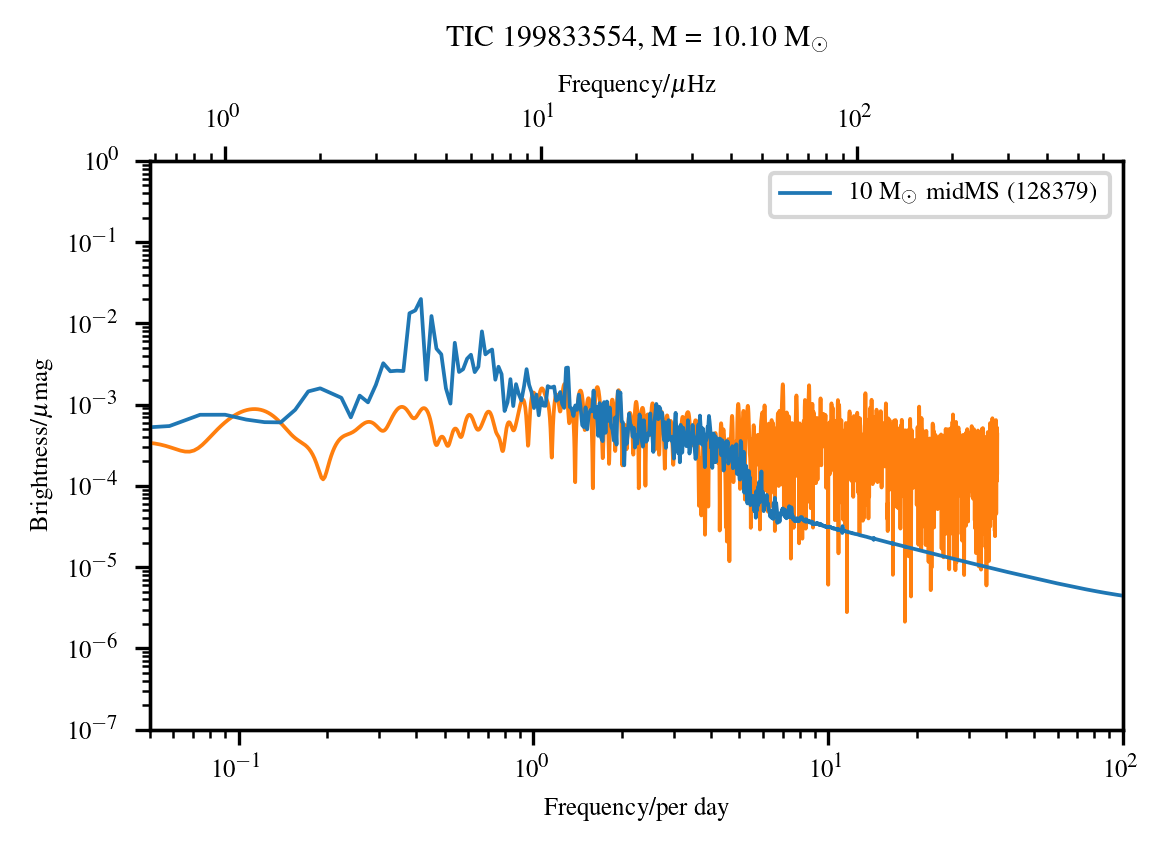}
        \caption{Normalised temperature perturbation (solid blue line) from a 10 \msol{} star simulation and brightness variation (solid orange
line) from a 10.10 \msol{} star from the TESS catalogue as a function of wave frequencies.}
\end{figure}
\begin{figure}[ht!]
        \centering
        \includegraphics[trim={0.0cm 0.0cm 0 0.0cm},clip,width=0.9\columnwidth]{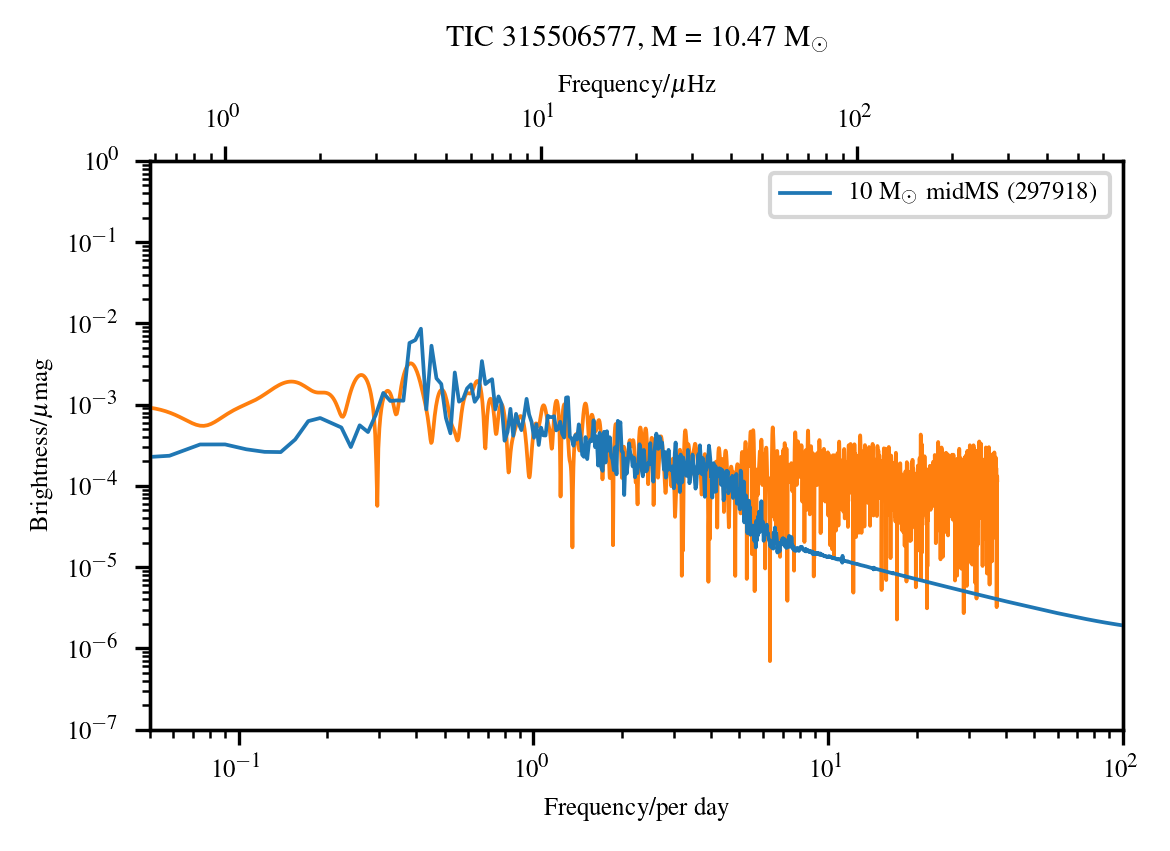}
        \caption{Normalised temperature perturbation (solid blue line) from a 10 \msol{} star simulation and brightness variation (solid orange
line) from a 10.47 \msol{} star from the TESS catalogue as a function of wave frequencies.}
\end{figure}
\begin{figure}[ht!]
        \centering
        \includegraphics[trim={0.0cm 0.0cm 0 0.0cm},clip,width=0.9\columnwidth]{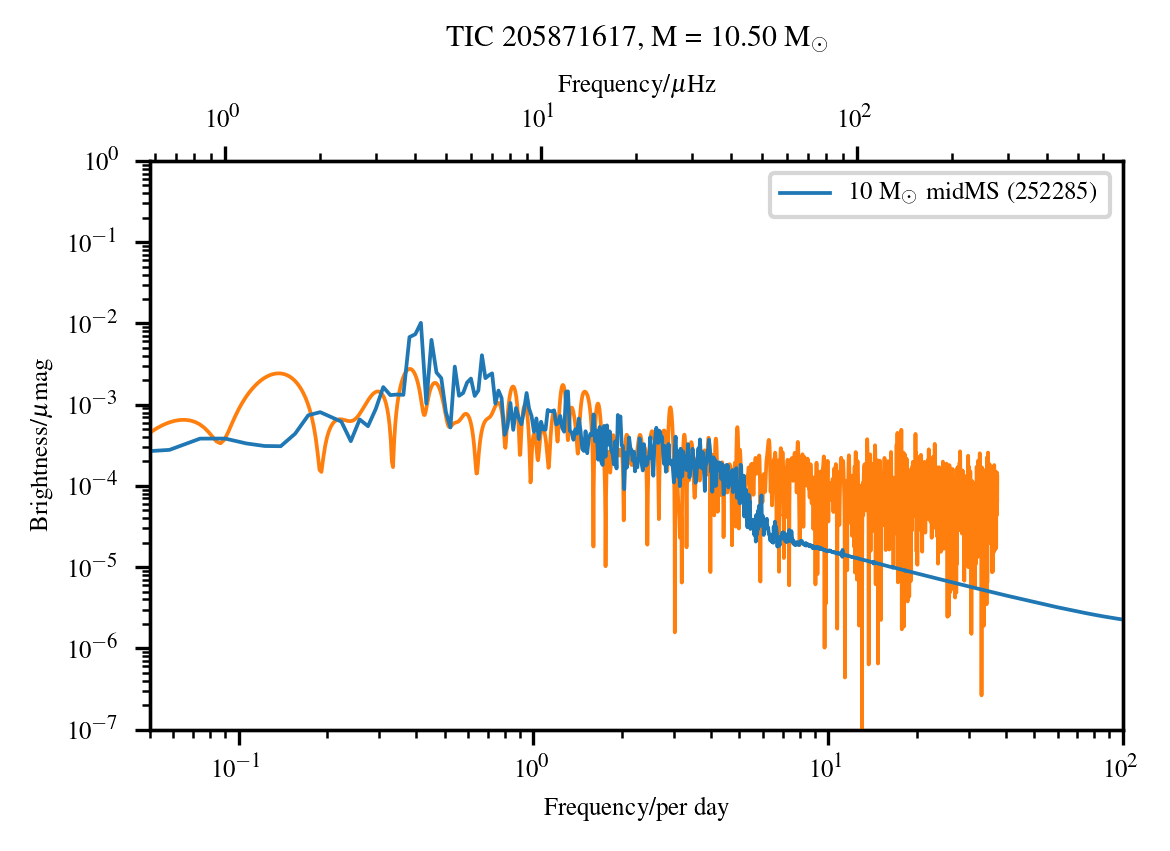}
        \caption{Normalised temperature perturbation (solid blue line) from a 10 \msol{} star simulation and brightness variation (solid orange
line) from a 10.50 \msol{} star from the TESS catalogue as a function of wave frequencies.}
\end{figure}
\begin{figure}[ht!]
        \centering
        \includegraphics[trim={0.0cm 0.0cm 0 0.0cm},clip,width=0.9\columnwidth]{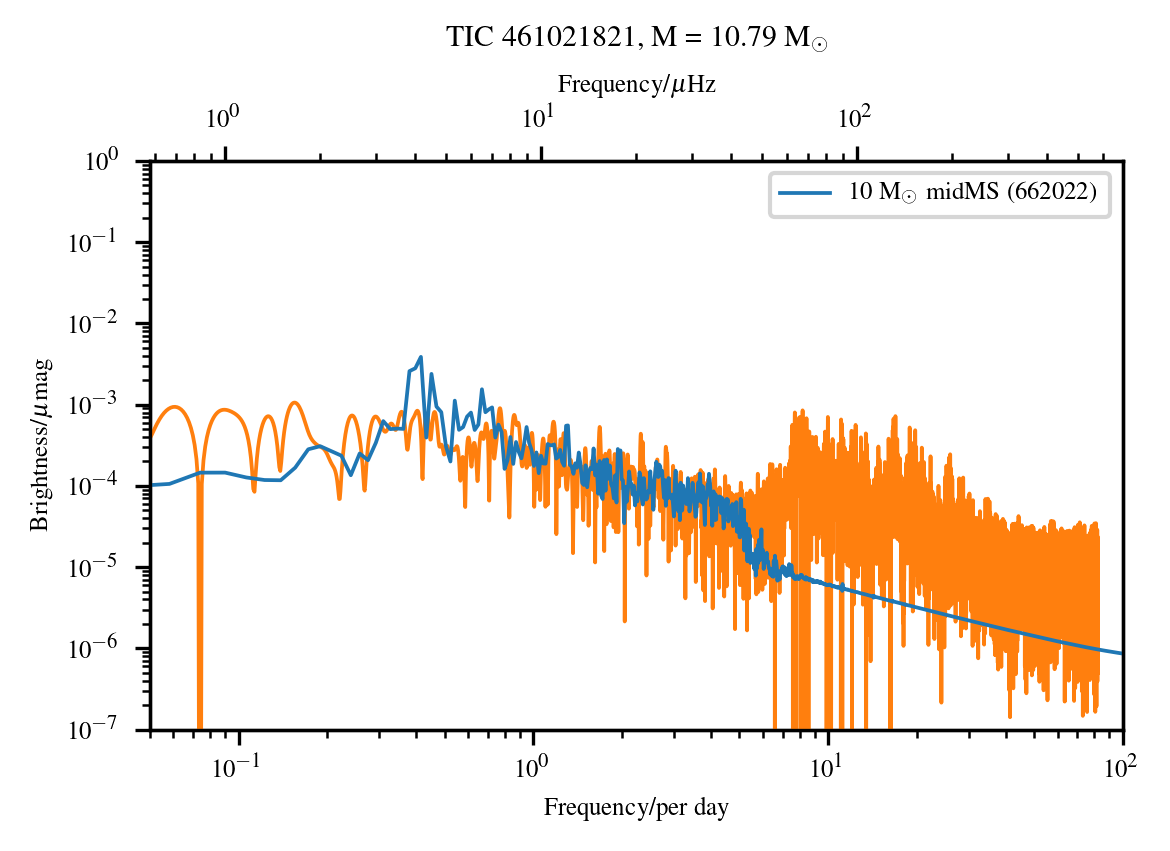}
        \caption{Normalised temperature perturbation (solid blue line) from a 10 \msol{} star simulation and brightness variation (solid orange
line) from a 10.79 \msol{} star from the TESS catalogue as a function of wave frequencies.}
\end{figure}
\begin{figure}[ht!]
        \centering
        \includegraphics[trim={0.0cm 0.0cm 0 0.0cm},clip,width=0.9\columnwidth]{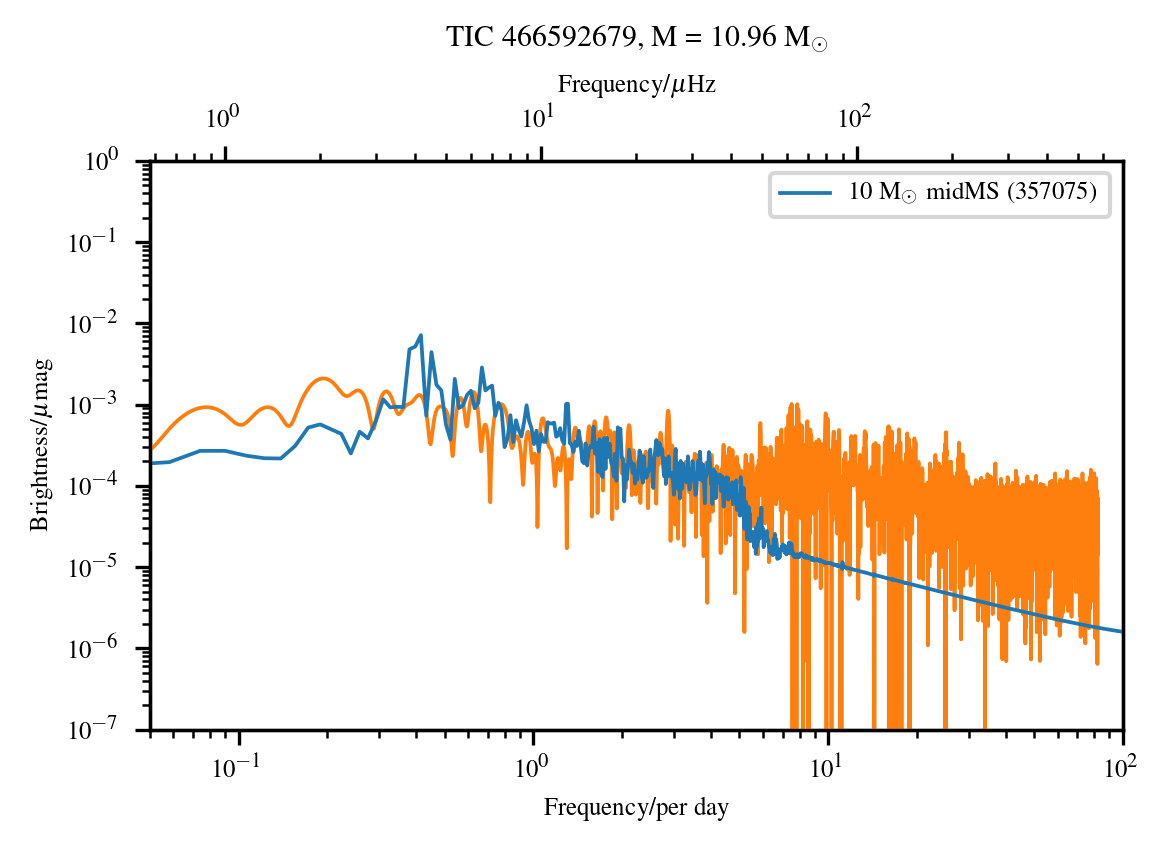}
        \caption{Normalised temperature perturbation (solid blue line) from a 10 \msol{} star simulation and brightness variation (solid orange
line) from a 10.96 \msol{} star from the TESS catalogue as a function of wave frequencies.}
\end{figure}

\begin{figure}[ht!]
        \centering
        \includegraphics[trim={0.0cm 0.0cm 0 0.0cm},clip,width=0.9\columnwidth]{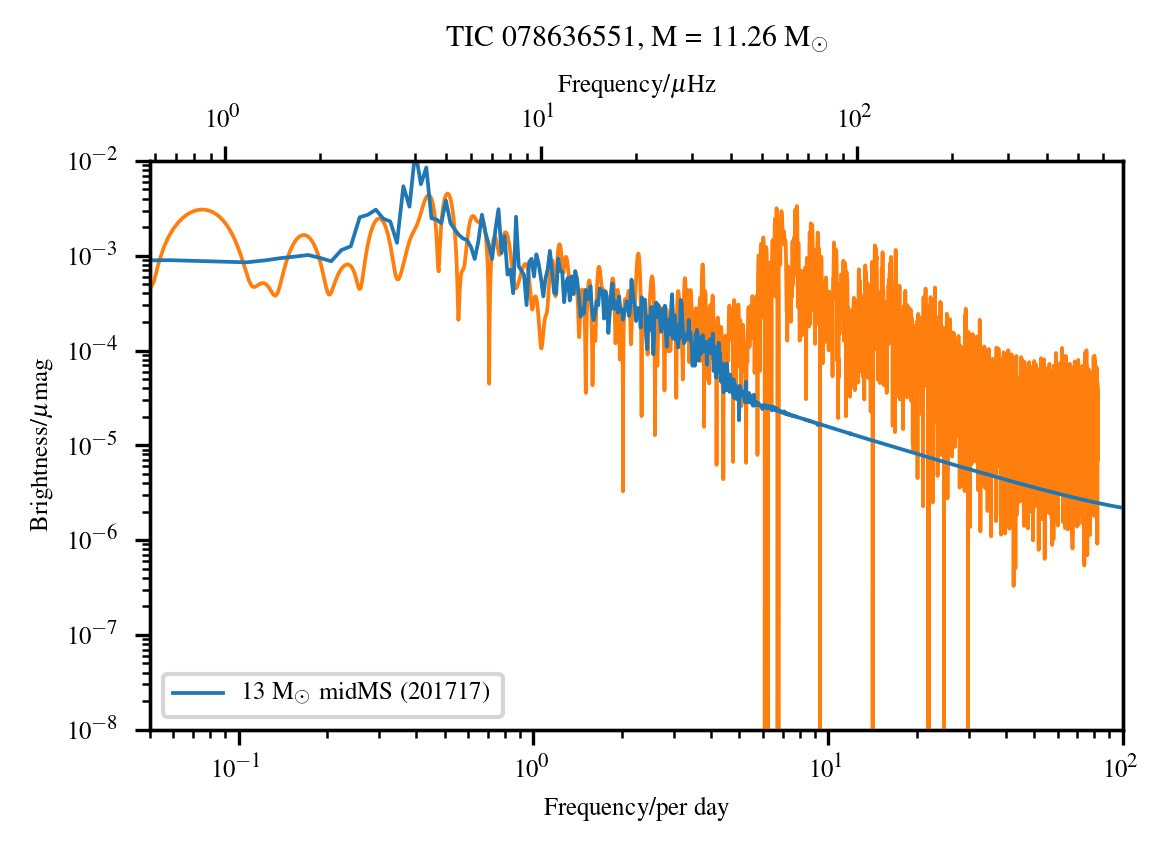}
        \caption{Normalised temperature perturbation (solid blue line) from a 13 \msol{} star simulation and brightness variation (solid orange
line) from a 11.26 \msol{} star from the TESS catalogue as a function of wave frequencies.}
\end{figure}
\begin{figure}[ht!]
        \centering
        \includegraphics[trim={0.0cm 0.0cm 0 0.0cm},clip,width=0.9\columnwidth]{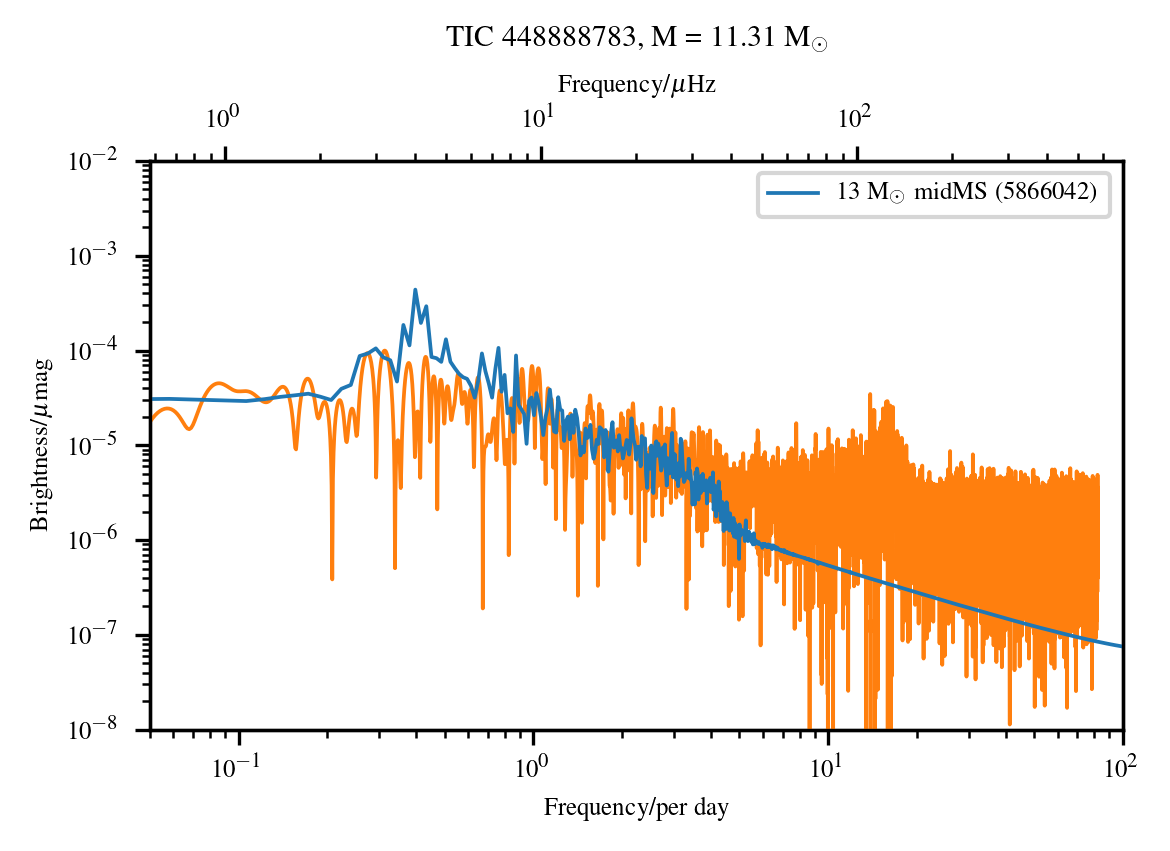}
        \caption{ Normalised temperature perturbation (solid blue line) from a 13 \msol{} star simulation and brightness variation (solid orange
line) from a 11.31 \msol{} star from the TESS catalogue as a function of wave frequencies.}
\end{figure}
\begin{figure}[ht!]
        \centering
        \includegraphics[trim={0.0cm 0.0cm 0 0.0cm},clip,width=0.9\columnwidth]{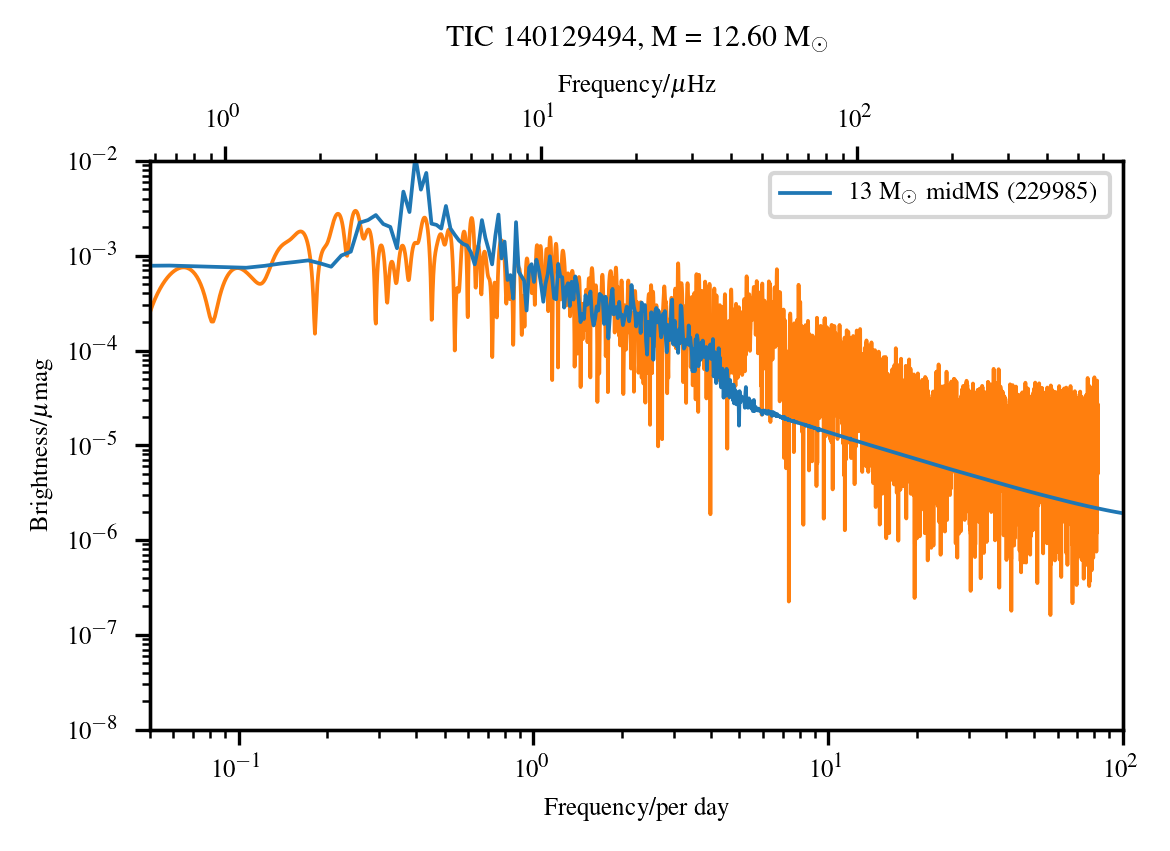}
        \caption{ Normalised temperature perturbation (solid blue line) from a 13 \msol{} star simulation and brightness variation (solid orange
line) from a 12.60 \msol{} star from the TESS catalogue as a function of wave frequencies.}
\end{figure}
\begin{figure}[ht!]
        \centering
        \includegraphics[trim={0.0cm 0.0cm 0 0.0cm},clip,width=0.9\columnwidth]{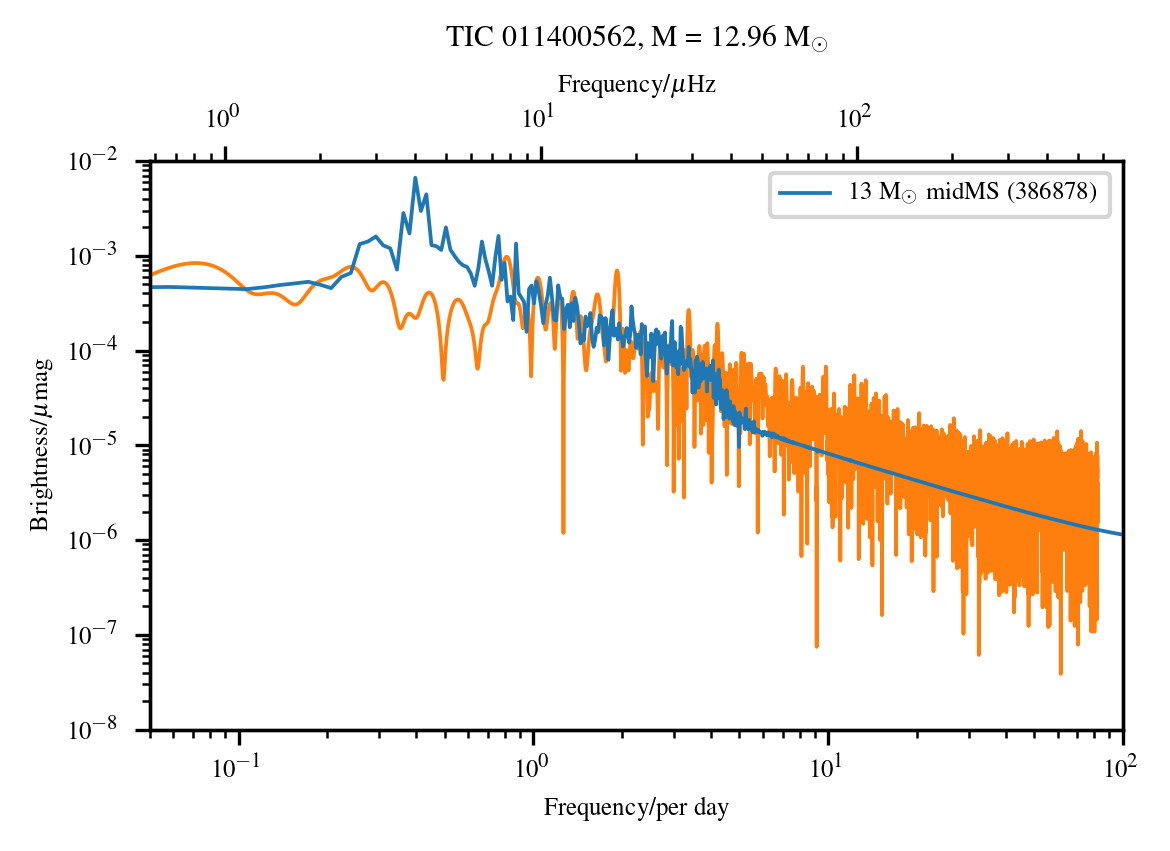}
         \caption{ Normalised temperature perturbation (solid blue line) from a 13 \msol{} star simulation and brightness variation (solid orange
line) from a 12.96 \msol{} star from the TESS catalogue as a function of wave frequencies.}
\end{figure}
\begin{figure}[ht!]
        \centering
        \includegraphics[trim={0.0cm 0.0cm 0 0.0cm},clip,width=0.9\columnwidth]{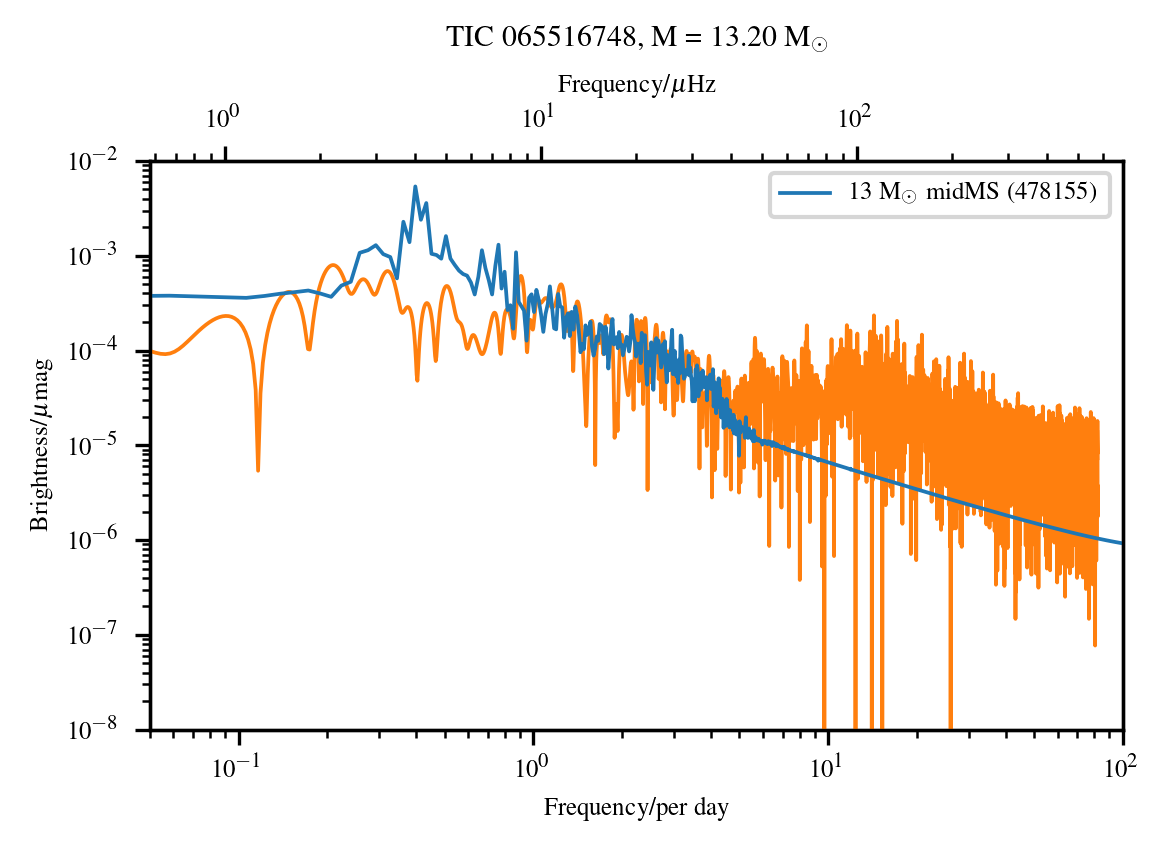}
        \caption{Normalised temperature perturbation (solid blue line) from a 13 \msol{} star simulation and brightness variation (solid orange
line) from a 13.20 \msol{} star from the TESS catalogue as a function of wave frequencies.}
\end{figure}
\end{appendix}

\end{document}